\documentclass[singlecolumn,aps,longbibliography]{revtex4-2}

\usepackage{amssymb}
\usepackage{amsmath}
\usepackage{tabularx}
\usepackage{multirow}
\usepackage[colorlinks]{hyperref}
\usepackage{soul,color}
\usepackage{graphicx}% Include figure files
\usepackage{dcolumn}% Align table columns on decimal point
\usepackage{bm}% bold math

\usepackage{footmisc}

\begin{document}

\title{Electromechanical memcapacitive neurons for energy-efficient spiking neural networks}



\author{Zixi Zhang}
\affiliation{School of Physics, Peking University, Beijing 100871, China}

\author{Yuriy~V.~Pershin}
\affiliation{Department of Physics and Astronomy, University of South Carolina, Columbia, SC 29208 USA}
\email{pershin@physics.sc.edu}

\author{Ivar~Martin}
\affiliation{Materials Science Division, Argonne National Laboratory, Argonne, IL 08540, USA}
\email{ivar@anl.gov}


%\date{\today}

\begin{abstract}
In this article, we introduce a new nanoscale electromechanical device -- a leaky memcapacitor -- and show that it may be useful for the hardware implementation of spiking neurons.  The leaky memcapacitor is a movable-plate capacitor that becomes quite conductive when the plates come close to each other. The equivalent circuit of the leaky memcapacitor involves a memcapacitive and memristive system connected in parallel. In the leaky memcapacitor, the resistance and capacitance depend on the same internal state variable, which is the displacement of the movable plate.  We have performed a comprehensive analysis showing that several spiking types observed in biological neurons can be implemented with the leaky memcapacitor. Significant attention is paid to the dynamic properties of the model.  As in leaky memcapacitors the capacitive and leaking resistive functionalities are implemented naturally within the same device structure, their use will simplify the creation of spiking neural networks.
\end{abstract}

%\keywords{}

\maketitle

\section{Introduction}

Information processing in biological systems relies on a complex network of interacting neurons. Each neuron, when subjected to a stimulus responds by outputting a signal that typically has a form of unharmonic spikes.
A variety of models of the spiking phenomenon have been proposed, most famously  the  Hodgkin–Huxley (HH) model, which successfully captured many observed features of spiking in neuronal membranes~\citep{hodgkin_currents_1952,hodgkin_components_1952,hodgkin_dual_1952,hodgkin_quantitative_1952,hodgkin_measurement_1952,izhikevich_dynamical_2007,gerstner_neuronal_2014}. This model prompted the development of bio-chemically based information processing models in the subsequent decades~\citep{catterall_hodgkin-huxley_2012,zhou_noise-induced_2003,kang_dynamic_2016,izhikevich_dynamical_2007}.

HH model attempts to accurately capture the properties of ion channels with memory and consequently is quite complex.  A number of simplified models, like the integrate-and-fire model, appeared  that can achieve a variety of spiking behaviors in response to different types of stimulation~\citep{abbott_lapicques_1999,fuortes_interpretation_1962,koch_methods_1998,gerstner_neuronal_2014,bryant_spike_1976,izhikevich_simple_2003,izhikevich_dynamical_2007}.
Despite being distinct from HH, one may expect that the qualitative features such as  instabilities/bifurcations, limit cycles, and synchronization among neurons are quite universal and robust  due to the general underlying principles of the dynamical systems~\citep{strogatz2018nonlinear}.
The simplified models, however, may have the important advantage of being easier to implement artificially using existing materials and devices. This leads to an exciting possibility of  biologically-inspired {\em neuromorphic} information processing systems implemented in the solid-state.

In the above context, the class of memory circuit elements~\citep{di_ventra_circuit_2009} becomes increasingly important because of the capacity of memory circuit elements to store and process the information on the same physical platform~\citep{diventra13a}. The memory circuit elements (in a pure form) are resistors, capacitors, and inductors with memory whose response is defined by the equations
\begin{eqnarray}
y(t)&=&g\left({\bf x},u \right)u(t) \label{eq:1a} \;\;,\\
\dot{{\bf x}}&=&{\bf f}\left({\bf x},u\right) \;\; , \label{eq:1b}
\end{eqnarray}
where  $y(t)$  and $u(t)$  are any two complementary circuit variables (i.e., current, charge, voltage, or flux), $g\left({\bf x},u \right)$ is a generalized response,
${\bf x}$ is a set of $n$ state variables describing the internal state of the device, and ${\bf f}\left({\bf x},u \right)$ is a continuous $n$-dimensional vector function.
Depending on the choice of the complementary circuit variables, Eqs.~(\ref{eq:1a}) and (\ref{eq:1b}) are used to define memristive, memcapacitive, or meminductive systems~\citep{di_ventra_circuit_2009}. Examples of physical realizations of memory circuit elements can be found in the review paper~\citep{pershin_memory_2011}. %Interestingly, the combination ...

For more than a decade, significant attention has been paid to the application of memristive systems in neuromorphic computing. Indeed, the memristive systems share several common characteristics with biological synapses such as the two-terminal structure, adaptivity, and high integration density. One of the first works  in this area was the demonstration of associative memory with memristive neural networks by one of us~\citep{pershin09c}.  Interestingly, the relevance of the memristive equations to the HH model was established already in 1976 by Chua and Kang ~\citep{chua76a,chua_hodgkinhuxley_2012}. Specifically, they proposed that ``the potassium channel of the Hodgkin-Huxley model should be identified as a first-order time-invariant voltage-controlled memristive one-port and the sodium channel should be identified as a second-order time-invariant voltage-controlled memristive one-port''~\citep{chua76a}. Even though most of the existing research focuses on purely electronic schemes, there are also models that rely on the mechanical realization of memory and spikes~\citep{heimburg_capacitance_2012,chen_computational_2019,jing_electric_2018,galassi_coupling_2021,holland_thinking_2019}.

Neuromorphic applications of memcapacitive systems have received much less attention.  This may be explained by the fact that in general memcapacitive devices~\citep{martinez2010solid,martinez2011bistable,najem_dynamical_2019,liu_new_2020} have been much less studied (in comparison to the memristive ones). At the same time, the reactive nature of the capacitive  response is very promising for low-power computing applications as at equilibrium the memcapacitive devices do not consume any power. Thus, currently, the application of memcapacitors in neuromorphic circuits~\citep{pershin2014memcapacitive,scott2022evidence} is a narrow but highly promising research field. In fact, the possibility of energy-efficient neuromorphic computing  with solid-state memcapacitive structures has been demonstrated recently~\citep{parkin21a}.

% The memcomputing is not directly related to the neuromorphic computing...

%The features common both to the models of neurons and neuromorphic devices  is the presence of an internal variable that controls some electrical properties (e.g. a conductance of an ion channel).  This is also a key ingredient of  {\em memcomputing}, which extends the conventional digital computing by endowing the logical elements with an internal state variable.
%\citep{pershin_memory_2011,liu_new_2020,bearden_instantons_2018,cadareanu_rebooting_2019,caravelli_complex_2017,manukian_accelerating_2019,strukov_missing_2008,pershin_demonstration_2019,traversa_polynomial-time_2017,traversa_application_2020,di_ventra_perspective_2018,di_ventra_topological_2017,pershin_memcapacitive_2014,pershin_memcomputing_2014,di_ventra_circuit_2009,pershin_practical_2010,traversa_universal_2015,ventra_physical_2013,peotta_superconducting_2014,pershin_memristive_2009,ascoli_synchronization_2015,pfeiffer_quantum_2016,najem_dynamical_2019}.  Already, memcomputing devices have shown promise in solving complex optimization problems
%\citep{manukian_accelerating_2019,pershin_memcapacitive_2014,pershin_memristive_2009,ascoli_synchronization_2015,ghosh-dastidar_spiking_2009}.

%Interestingly, both in memcomputing devices and in neurons, the canonical internal state variable corresponds to electric charge: in HH model, it is the ionic concentration that controls the ion channel conductance, while in memristors, the resistivity is a function of the charge that has flown through the device [LChua].

The mechanical aspect of biological neurons is also becoming increasingly recognized as important in the study of neuroscience.  They play a role in several physiological processes, including the generation of action potentials~\cite{ling2020high}.
 The mechanical changes in the cell membrane can influence the opening and closing of ion channels, which can impact the electrical signaling of the neuron.
The soliton model combines mechanical  and electrical factors of signal propagation through axons and has been able to account for some experimental effects in anesthesiology beyond the Hodgkin–Huxley model~\citep{johnson_soliton_2018,appali_comparison_2012,andersen_towards_2009,heimburg_thermodynamics_2007}.

In this work, in part inspired by  biological neurons, we propose a simple neuromorphic electromechanical model based on a leaky memcapacitor that is capable of  achieving the periodic generation of spikes under constant stimulation.
The presence of  additional ---mechanical--- degrees of freedom makes it easier to realize neuromorphic behaviors in artificial electromechanical systems.
Applying the methods of dynamical systems, we explore the qualitative features of this model in terms of its fixed points, bifurcations, and limit cycles under DC stimulation. In addition, we demonstrate that the system is capable of  complex dynamical adaptations, such as synchronization with periodic external drives, spiking frequency drift, bursting, and other adaptations (for some behaviors, an additional memory circuit element is required).

\section{The model}
\begin{figure}[t]
  \centering
(a)  \includegraphics[height=4.5cm]{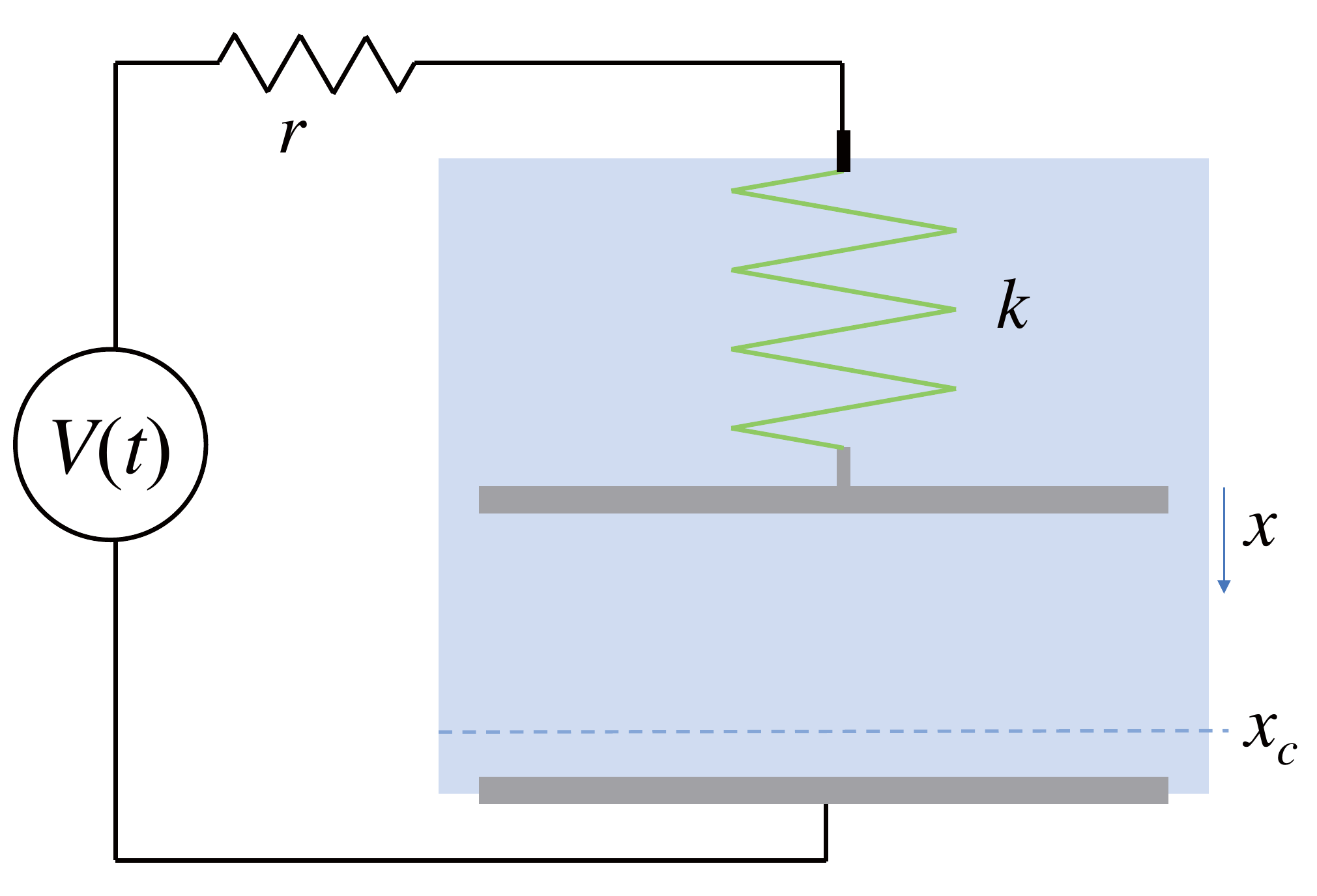} \;\; \;\;\;\;
(b) \includegraphics[height=4.3cm]{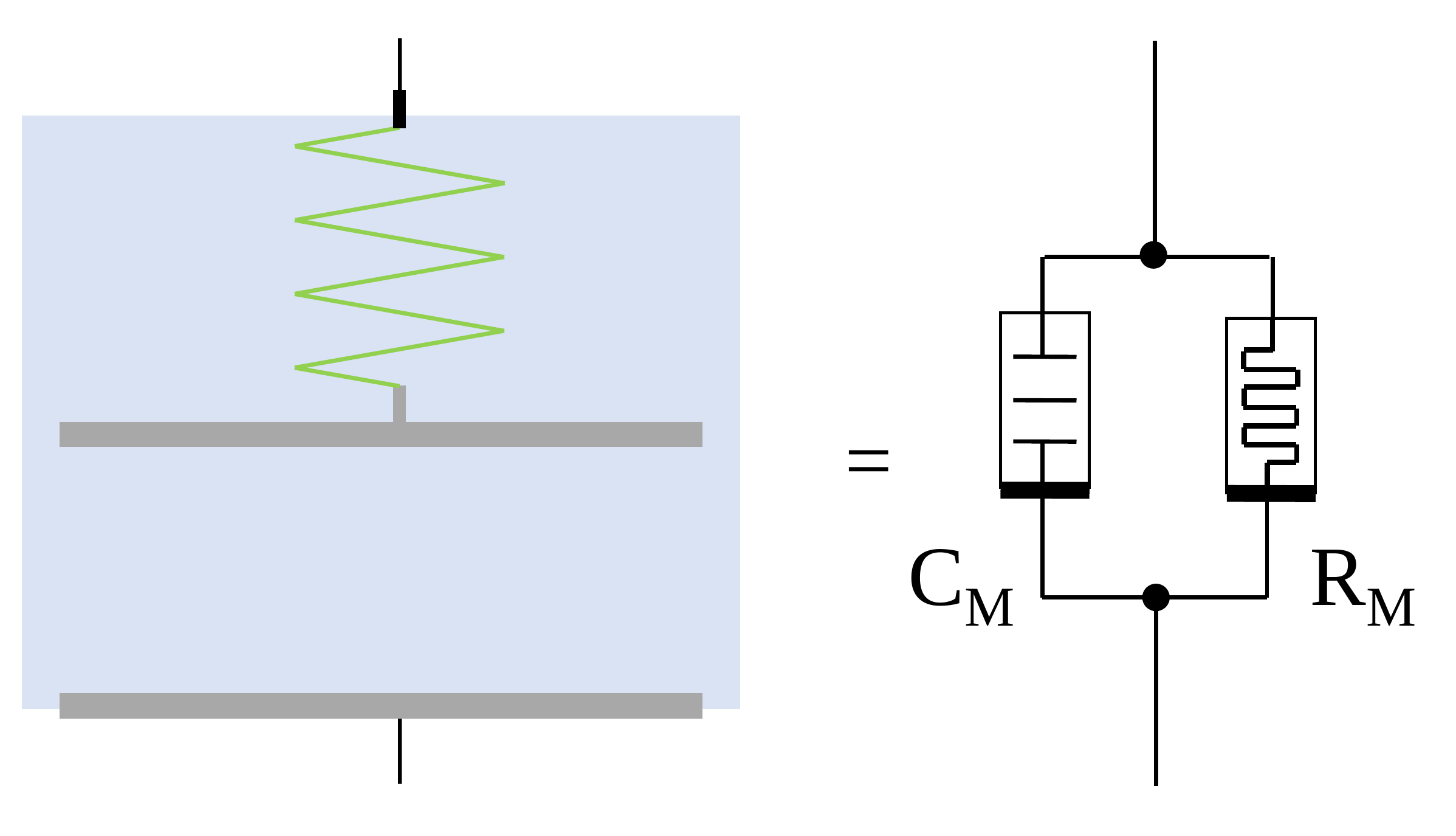}
    \caption{%
  (a) An electromechanical leaky memcapacitor connected to a voltage source through a resistor. (b) The equivalent electronic circuit of the leaky memcapacitor: a memcapacitive system, C$_\textnormal{M}$, and memristive system, R$_\textnormal{M}$, connected in parallel. Both systems depend on the same internal state variable $x$.
  }
  \label{fig:1}
\end{figure}

%We are devising a system with periodic electronic spikes to simulate neuron membranes with a leaky memcapacitor.
In this section, we introduce a leaky memcapacitor, its model, and the circuit we use to simulate the spiking neurons. The circuit, physical diagram of leaky memcapacitor, and its equivalent circuit  are shown in Fig.~\ref{fig:1}.

The central element of our spiking neuron is a {\em leaky memcapacitor} -- a capacitor with a plate that moves in  response to the force exerted by the internal electric field and restoring force of the spring (the spring constant is $k$).  In the absence of charges on the plates, $q = 0$, the distance between two plates is $d$. It is assumed that the displacement of the top plate, $x$, is positive
for the displacement towards the bottom plate, see Fig.~\ref{fig:1}(a). Thus, the distance between the plates is $d-x$.

We assume that the capacitor is {\em leaky}: there is a finite  resistance $R(x)$, which depends on the distance between the plates, dropping rapidly in the `contact' region, $x > x_c$. The leaky memcapacitor has the {\em memory} (which justifies the term memcapacitor) because the current position of the moving plate (and hence the capacitance itself) depends on the  prior history of the charge on the capacitor, which in turn depends on the history of applied voltage.
It is this history dependence, the memory, that is responsible for the emergence of complex behaviors that the simple circuit in Fig.\ref{fig:1}(a) exhibits.

The main operating principle of this device is straightforward. Starting from the equilibrium position at $x=0$, upon turning on voltage $V$, the capacitor begins to charge up. This leads to the attraction between its plates, which brings them closer together. For large enough applied voltage, the top plate reaches the contact region, the capacitor discharges, and the plate recoils back from the contact region. Note that the process of charging is limited by the resistor with the resistance $r$ (Fig.~\ref{fig:1}(a)).

%To avoid the plates finding a new static equilibrium position in the contact region, we found that it is helpful to  assume that the net restoring force drops in that region, which may be interpreted as a proximity-induced attraction between the two surfaces

We found that the proximity-induced attraction between the two surfaces (plates) is helpful to avoid the plates finding a new static equilibrium position in the contact region.  We model this part of interaction with a Lennard-Jones-like potential. The short-range attraction between the plates allows for a more effective discharge process, leading to a periodic  approach of the top plate to the bottom plate followed by recoil. The exact form of $R(x)$ and the form of the potential, to a certain extent, do not matter.  In the biological context, our memcapacitor may represent two lipid monolayers forming a cell membrane. As the membrane swells or thins, the resistance of the membrane (via ion channels) is affected, as modeled by $R(x)$.

In our modeling of membrane dynamics, we neglect the kinematic mass (in a biological setting that would be because membranes reside in an aqueous environment that has  considerable viscosity). This overdamped regime makes it nontrivial to have oscillatory behavior, and new types of mechanisms are required to perform the periodic spiking that we see possible.

Eqs. (\ref{eq:1a}) and (\ref{eq:1b}) provide the general framework for the description of leaky memcapacitor. For the capacitive and resistive responses, Eq. (\ref{eq:1a}) is written as
\begin{eqnarray}
        q&=&\frac{\epsilon A}{d-x}V_C\equiv C(x) \cdot V_C \label{eq:MemC} \;\;,  \\
        V_C&=&\left[ R_m\cdot\left(\frac{1}{\pi}\text{arctan}\beta\left(x_c-x\right)+0.5\right)+\rho_0\frac{d-x}{A} \right]I_M\equiv R(x) \cdot I_M \;\;,
        \label{eq:MemR}
\end{eqnarray}
where $q$ is the memcapacitor charge, $V_C$ is the voltage across the leaky memcapacitor, $I_M$ is the leakage current, $\varepsilon$ is the permittivity, $A$ is the plate area, and $R_m$, $\beta$, $\rho_0$ are the parameters defining the memristance. In the normal regime, $R(x)\approx R_m$, while in the contact regime, $R(x)$ is mainly determined by the $\rho_0$ term; $\beta$ describes the  rate of change of the $R_m$ component in $R(x)$ in the vicinity of $x_c$.

\begin{figure}[tb]
  \centering
(a)  \includegraphics[width=7.5cm]{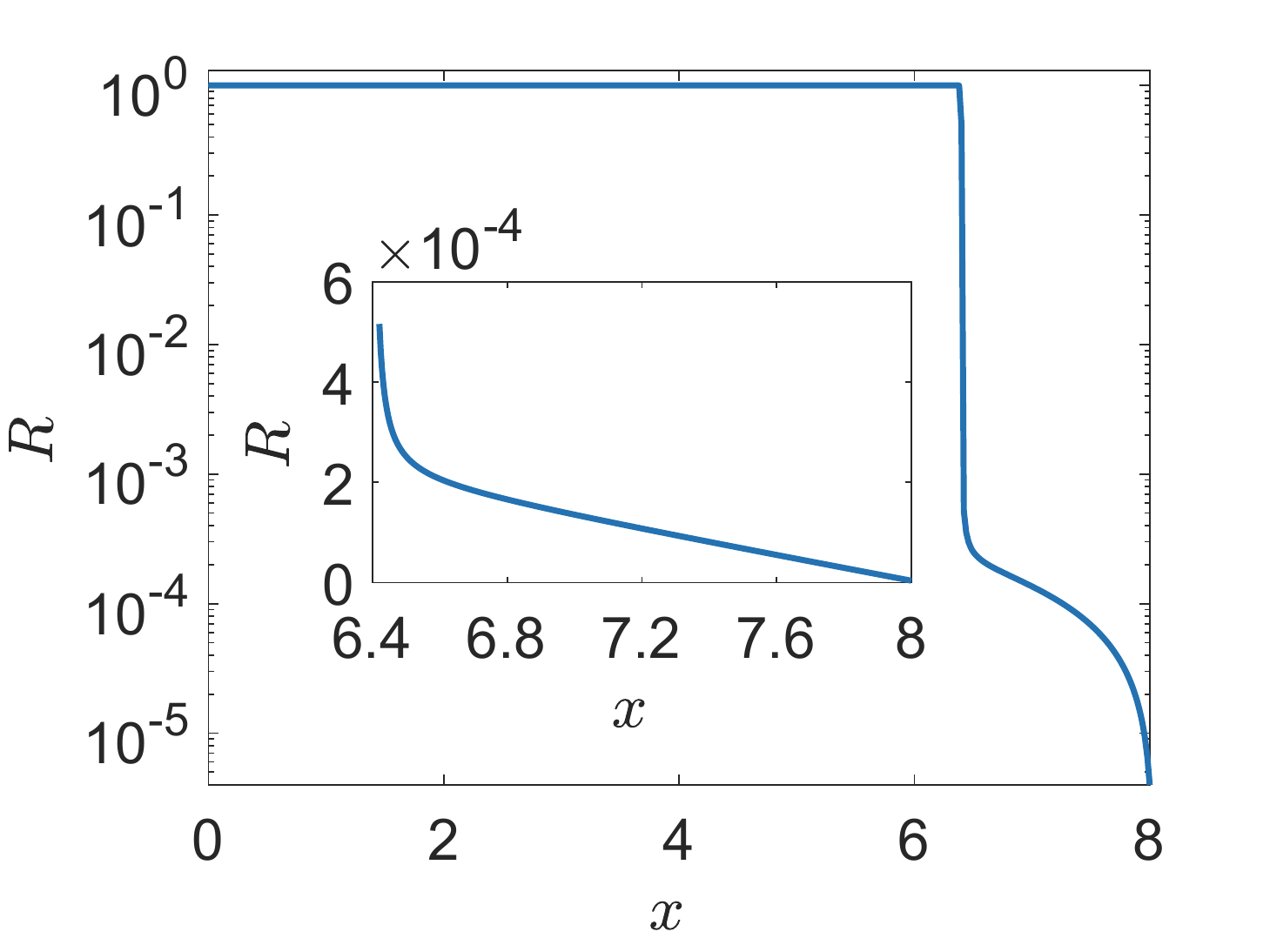} \;\; \;\;\;\;
(b) \includegraphics[width=7.5cm]{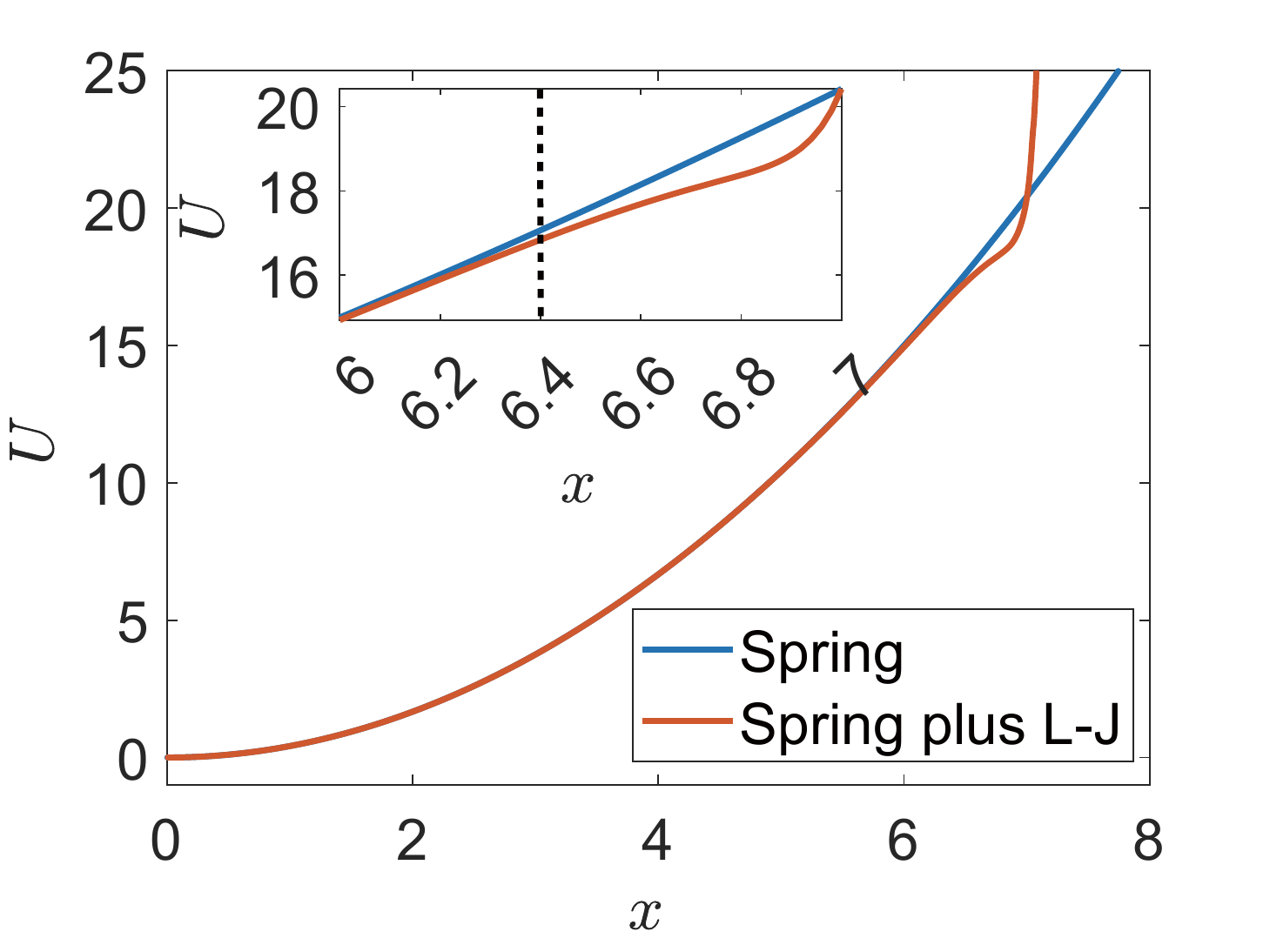}
    \caption{%
    (a) Resistance as a function of the displacement of the top plate (Eq.~(\ref{eq:MemR})). (b) Potential energy as a function of the displacement of the top membrane (Eq.~(\ref{eq:pot})).
  The insets present a zoomed-in contact region. The dashed line refers to $x_c$.
  }
  \label{fig:LJR}
\end{figure}

The leaky memcapacitor is described by a single internal state variable $x$, which is the displacement of the top plate from its equilibrium position (at $q=0$) in the downward direction (see Fig.~\ref{fig:1}(a)). Its dynamics (corresponding to Eq.~(\ref{eq:1b})) is represented by
\begin{equation}
    \gamma \dot{x}=\frac{q^2}{\epsilon A}-\frac{\textnormal{d} U(x)}{\textnormal{d} x} \;\;, \label{eq:x}
\end{equation}
where $\gamma$ is the dissipation coefficient, and the potential $U(x)$ is chosen as
\begin{equation}
    U(x)=\frac12 kx^2 + 4\epsilon_l\left[\left(\frac{\sigma}{d-x}\right)^{12}-\left(\frac{\sigma}{d-x}\right)^6\right]  \label{eq:pot} \;\;.
\end{equation}
Here, the first term is the spring potential energy, while the second one is the Lennard-Jones-like potential that we use to describe the contact interaction between the plates. In Eq.~(\ref{eq:pot}), $k$ is the spring constant, $\epsilon_l$ is the depth of the Lennard-Jones  potential well, and $\sigma$ is the distance at which the Lennard-Jones potential energy is zero.

%The behaviors of the whole potential and the resistance are shown in Fig. \ref{fig:LJR}. From a physical perspective, we may expect the resistance $R$ proportional to the separation $d-x$ in the normal regime, while a constant one is more ordinary in mechanical devices. In Appendix A we also provide some comparisons of analyses based on the two kinds, where we find the difference is quite small.

In the following, we measure the distances in the units of $\sigma$, resistances in the units of $R_m$, $U$ in the units of $\epsilon_l$, $k$ in the units of $\epsilon_l/\sigma^2$, time in the units of $R_m A\epsilon/\sigma$, charge in the units of $\sqrt{\epsilon\epsilon_lA/\sigma}$, $\gamma$ in the units of $\epsilon\epsilon_lAR_m/\sigma^3$, voltage in the units of $\sigma\epsilon_l/\epsilon A$, and current is the units of $\sqrt{\sigma\epsilon_l/\epsilon A}/R_m$. For thus defined dimensionless variables and parameters, we use the original notation to minimize clutter in the text.% and define them by stars on the figures.

%using dimensionless variables defined as $x/\sigma\to x$, $\tilde d=d/\sigma$, $r^*=r/R_m$, $R^*(x)=R(x)/R_m$, $\beta^*=\beta\sigma$, $\rho_0^*=\rho_0(\sigma/R_mA)$, $U^*=U/\epsilon_l$, $k^*=k(\sigma^2/\epsilon_l)$, $t^*=t(\sigma/R_m\epsilon)$, $q^*=q\sqrt{\sigma/\epsilon\epsilon_lA}$, $\gamma^*=\gamma(\sigma^3/\epsilon\epsilon_lAR_m)$, $V^*=V\sqrt{\epsilon A/\sigma\epsilon_l}$, and $I_M^*=I_M(R_m\sqrt{\epsilon A/\sigma\epsilon_l})$ %\textcolor{orange}{What character should we use for the new way of notation? like defining $V^*=1/\sqrt{\epsilon A/\sigma\epsilon_l}$ for $V$, $t^*=1/(\sigma/R_m\epsilon)$, $q^*=1/\sqrt{\sigma/\epsilon\epsilon_lA}$, $I^*=V^*/R_m$,... to simplify those with complex forms.}.

%The results presented below were obtained using dimensionless variables defined as $x/\sigma\to x$, $\tilde d=d/\sigma$, $r^*=r/R_m$, $R^*(x)=R(x)/R_m$, $\beta^*=\beta\sigma$, $\rho_0^*=\rho_0(\sigma/R_mA)$, $U^*=U/\epsilon_l$, $k^*=k(\sigma^2/\epsilon_l)$, $t^*=t(\sigma/R_m\epsilon)$, $q^*=q\sqrt{\sigma/\epsilon\epsilon_lA}$, $\gamma^*=\gamma(\sigma^3/\epsilon\epsilon_lAR_m)$, $V^*=V\sqrt{\epsilon A/\sigma\epsilon_l}$, and $I_M^*=I_M(R_m\sqrt{\epsilon A/\sigma\epsilon_l})$ %\textcolor{orange}{What character should we use for the new way of notation? like defining $V^*=1/\sqrt{\epsilon A/\sigma\epsilon_l}$ for $V$, $t^*=1/(\sigma/R_m\epsilon)$, $q^*=1/\sqrt{\sigma/\epsilon\epsilon_lA}$, $I^*=V^*/R_m$,... to simplify those with complex forms.}.
 The parameters used in our simulations are given in Table \ref{paramters}. Fig.~\ref{fig:LJR}(a) shows $R(x)$ as defined by Eq.~(\ref{eq:MemR});  Eq.~(\ref{eq:pot}) is presented in Fig.~\ref{fig:LJR}(b).

 Note that qualitatively similar results (to the ones presented in this paper) may be obtained using a different choice of parameters and functional dependencies. In particular, we have verified that the results remain nearly the same when the constant resistance for $x\lesssim 6$ in Fig.~\ref{fig:LJR}(a) is replaced with a resistance linearly dependent on $x$ (for more details, see the Supplemental Information (SI) Appendix A).

\begin{table}[!ht]
    \centering
    \caption{Parameters used in simulations.}
    \label{paramters}
    \begin{tabular}{cccc}
    \hline
    Parameter  & Value                              & Parameter    & Value               \\ \hline
    $d$        & $8$                               & $r$          & $10^{-3}$           \\
    $x_c$        & $6.4$           & $\rho_0$   & $1.25\times10^{-4}$              \\
    $\beta$      & $5\times10^4$         & $\gamma$   & $1.25\times10^{-4}$      \\
    $k$        & $5/6$                  &   &   \\
    \hline
    \end{tabular}
\end{table}

% \begin{table}[!ht]
%     \centering
%     \caption{Parameters used in simulations\textcolor{orange}{and table like this?}}
%     \label{paramters}
%     \begin{tabular}{cccc}
%     \hline
%     Parameter  & Value                              & Parameter    & Value               \\ \hline
%     $d/\sigma$        & $8$                               & $r/R_m$          & $10^{-3}$           \\
%     $x_c/\sigma$        & $6.4$           & $\rho_0/\rho_0^*$   & $1.25\times10^{-4}$              \\
%     $\beta\sigma$      & $5\times10^4$         & $\gamma/\gamma^*$   & $1.25\times10^{-4}$      \\
%     $k/k^*$        & $5/6$                  &   &   \\
%     \hline
%     \end{tabular}
% \end{table}

To simulate  Fig.~\ref{fig:1}(a) circuit, we use Kirchhoff's voltage law
\begin{equation}
    V(t)=r\left(\dot{q}+I_M \right)+V_C\;\; , \label{eq:K}
\end{equation}
which is applied together with the equations defining the leaky memcapacitor, Eqs.~(\ref{eq:MemC})-(\ref{eq:pot}). The trajectories $(x(t),q(t))$
are found by numerical integration of Eqs. (\ref{eq:x}) and (\ref{eq:K}). The integration was performed using the ODE solver ode45 for nonstiff differential equations in MATLAB (version R2022a).%, except for the one of the second type of memory effect in Section 5, which was performed using ode4 (Runge-Kutta) solver in Simulink$^\circledR$. %\textcolor{orange}{in that case, ode45 easily gets stuck. And should we also mention that the figures are plotted using MATLAB? (and add a reference of MATLAB here?)} \hl{YP: I think it is fine as it is now.}

%Below, all numerical values are given in dimensionless units. For the sake of clarity, we use the same notation for variables and parameters in dimensional and dimensionless units.

\section{Spiking behavior and phase diagram}
%\textcolor{red}{I think we should get rid of stars; just say that we are dealing with the rescaled variables per Table 1} \textcolor{orange}{This means we should define extra variables like $v=V^*/V=\sqrt{\epsilon A/\sigma\epsilon_l}$?} \hl{I would avoid the introduction of new variables. The last two new sentences in Sec. 2 may work. What do you think?}

\begin{figure*}[h]
  \centering
(a)  \includegraphics[width=5cm]{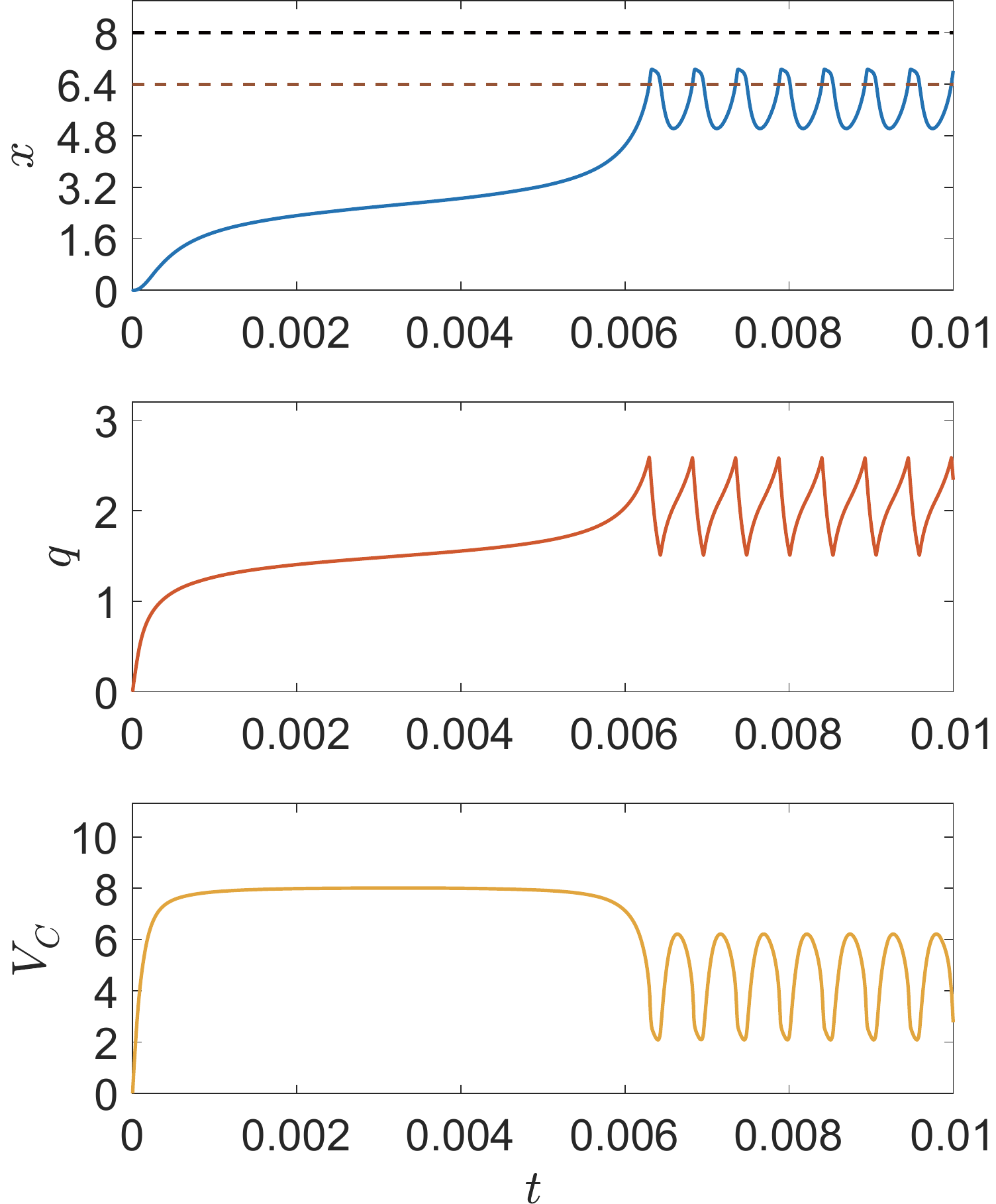}
(b)  \includegraphics[width=5cm]{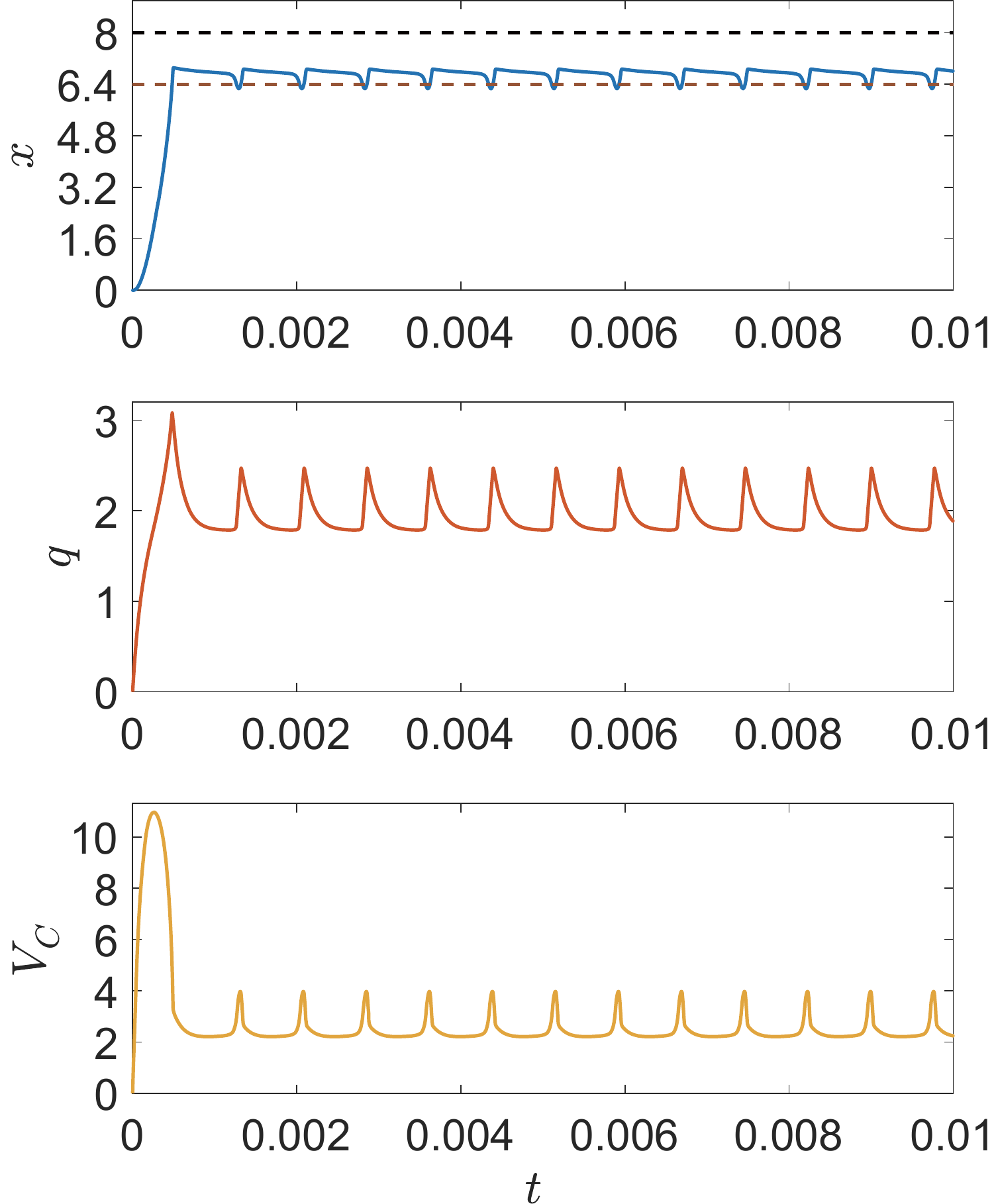}
(c) \includegraphics[width=5cm]{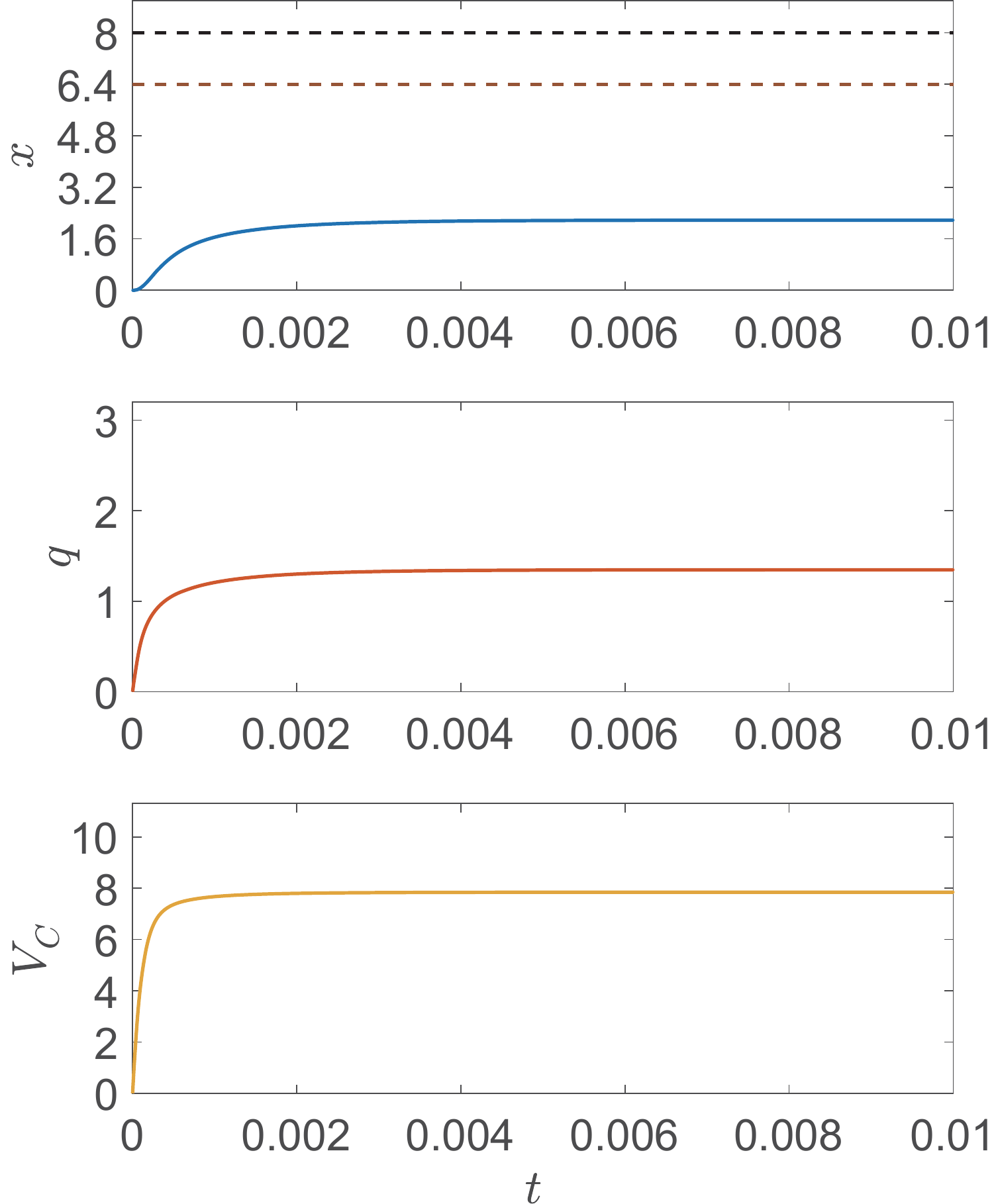}
    \caption{%
  The response to a step-like voltage applied  at $t=0$: starting from zero initial conditions ($x=0$, $q=0$), the circuit transits to (a), (b) a periodic spiking regime or (c) static regime. These plots were obtained using (a) $V=8.0829$, (b) $V=15.0111$, and (c) $V=7.8520$. In the top pannels, The black dashed line refers to $d$ and the brown one refers to $x_c$.
  }
  \label{fig:spike}
\end{figure*}

Figs.~\ref{fig:spike}(a) and (b) show two selected simulation results that demonstrate the transient dynamics from zero initial conditions, $x_0=0$ and $q_0=0$, to the regime of periodic spiking. These results indicate a significant dependence of the spike shape  on the magnitude of applied voltage ($V=8.0829$ in Fig.~\ref{fig:spike}(a) and $V=15.0111$ in Fig.~\ref{fig:spike}(b)).
According to Figs.~\ref{fig:spike}(a) and (b), the spikes are smoother at $V=8.0829$, and sharper at $V=15.0111$. The sharper spikes show a close resemblance to biological spikes.

A notable feature of the transient dynamics is the initial sharp increase in the voltage across the leaky memcapacitor, $V_C$. This property (clearly observed in Figs.~\ref{fig:spike}(a) and (b)) is associated with the smaller capacitance at short times (close to $t=0$) due to the initially large plate separation. Within the transient region, the leaky memcapacitor adapts to the applied voltage:
the Coulomb attraction reduces the distance between the plates what increases the capacitance.
The voltage plateau in Fig.~\ref{fig:spike}(a) is
close to the bifurcation point at $V_1'=7.9582$ (for more information, see SI Appendix A). This explains the relatively long duration of the transient region.

Moreover, in Fig.~\ref{fig:spike}(c) we present an example of non-spiking behavior. In this case, the trajectory ends in a sink (attractor).

\begin{figure*}[ht]
  \centering
  (a)\includegraphics[width=0.9\textwidth]{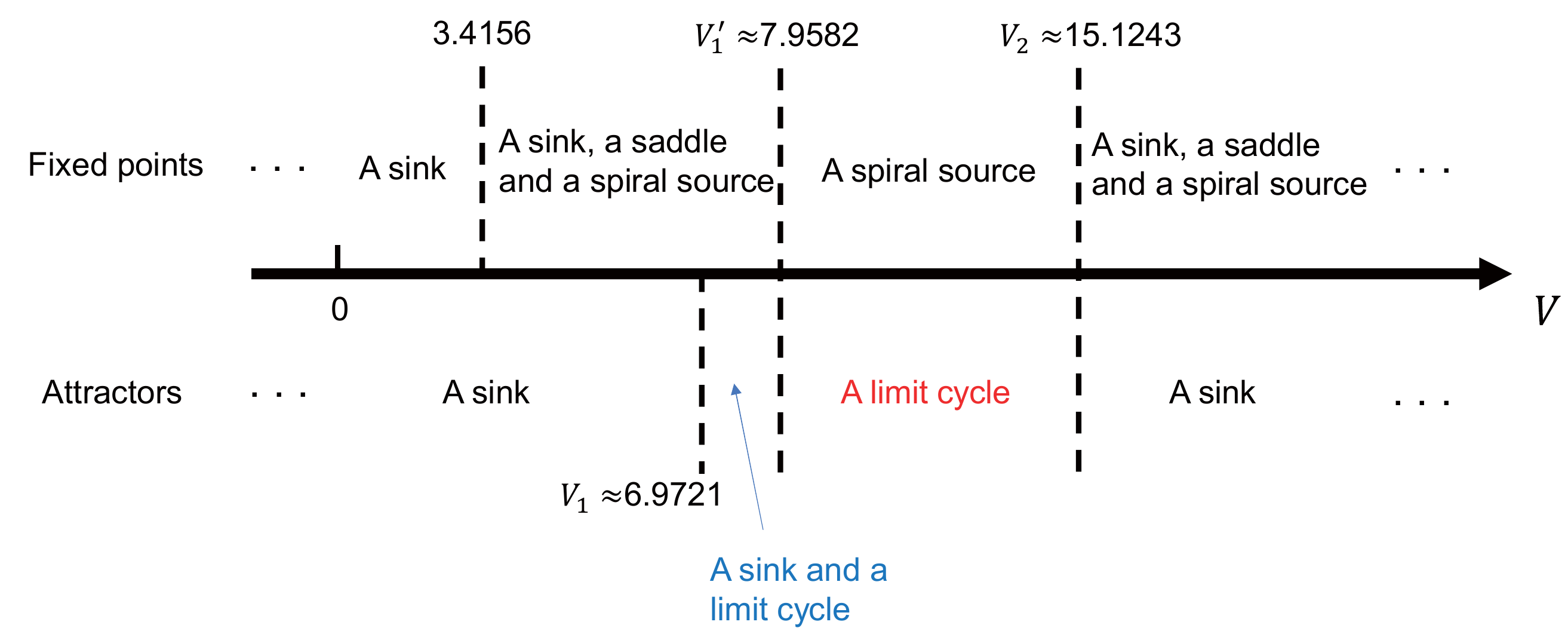}\\
(b)\includegraphics[width=5.1cm]{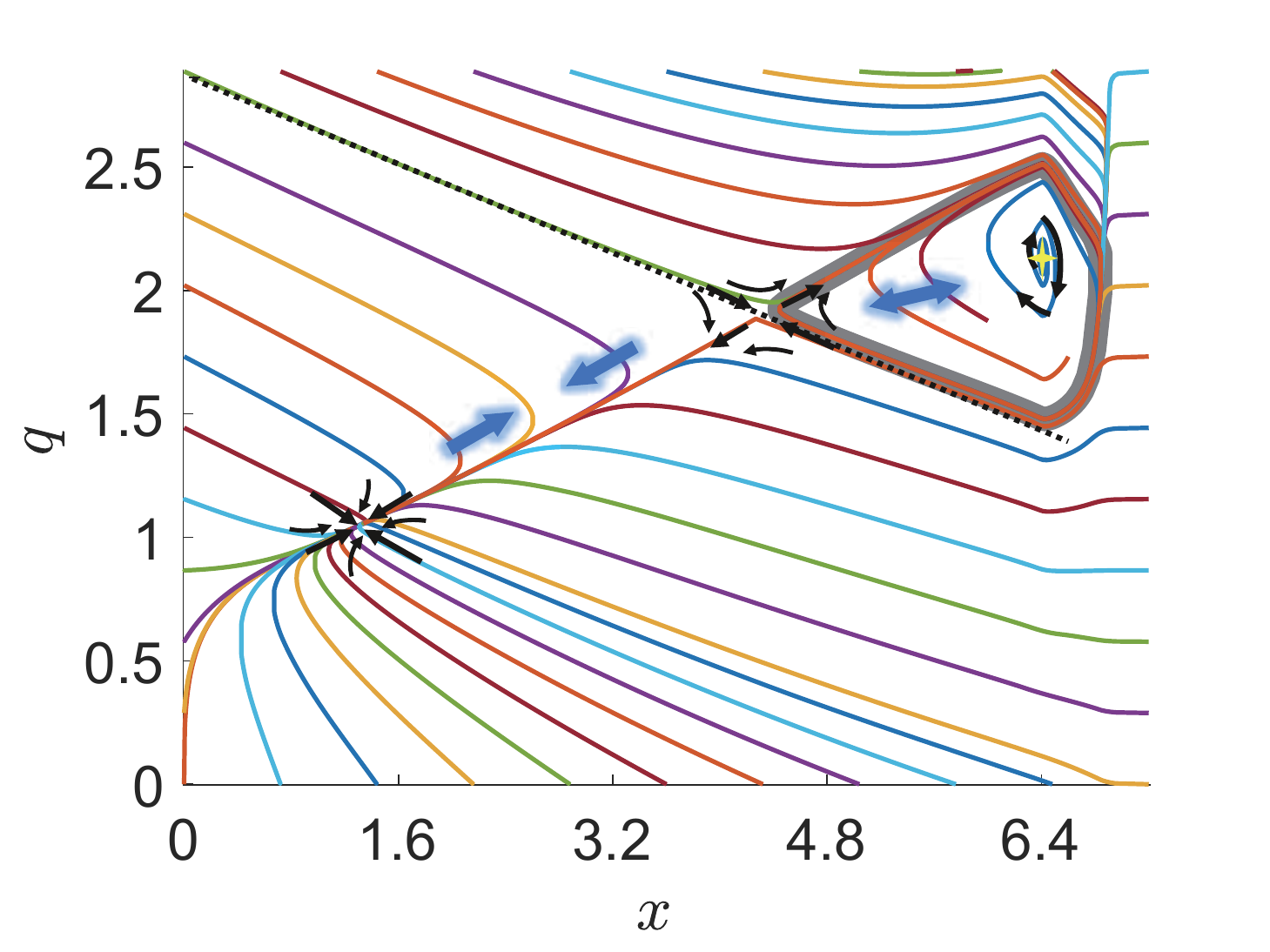}
(c)\includegraphics[width=5.1cm]{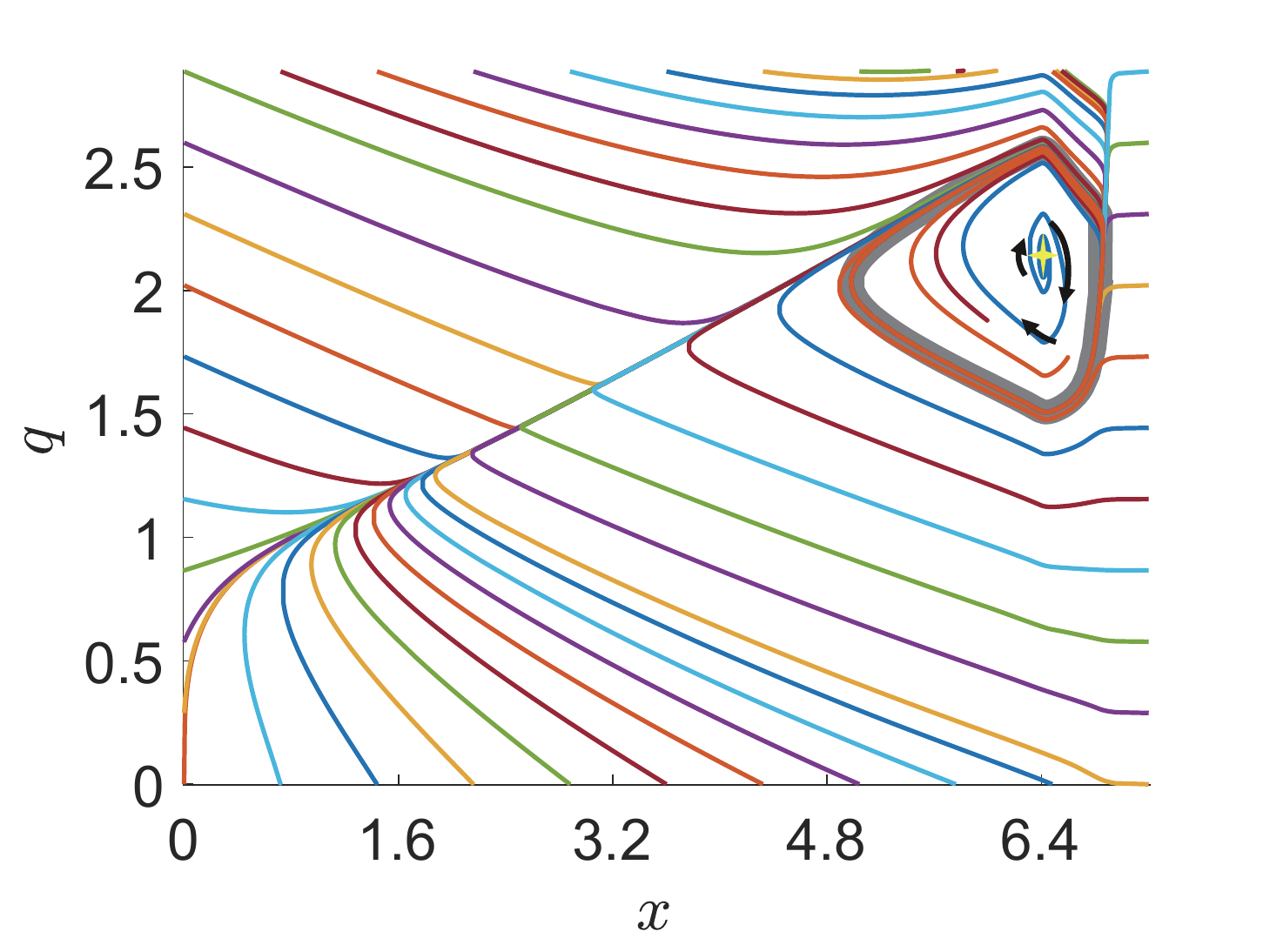}
(d)\includegraphics[width=5.6cm]{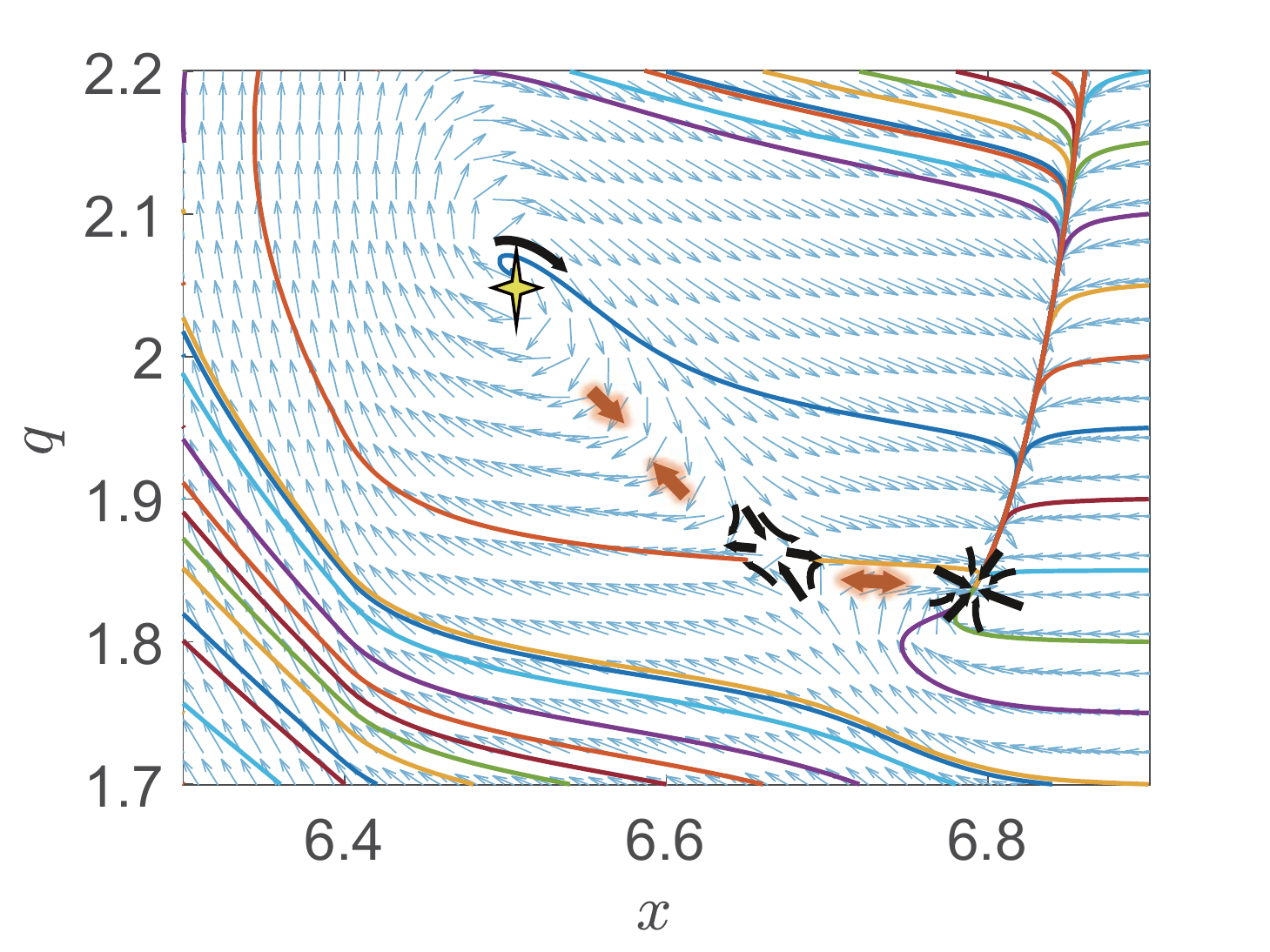}
  \caption{(a) An overall picture: the fixed points and attractors depending on the applied  voltage $V$.  The phase diagrams at (b) $V=7.0437$ (the sink and limit cycle case in (a)), (c) $V=7.9674$ (the limit cycle case in (a)) and (d) $V=15.5000$ (the sink case on the right of the limit cycle one in (a)). In (b) the phase space is divided into two parts by the flow lines towards the saddle, as depicted with a dashed line. The semi-transparent gray thick line represents the limit cycle. The saddle, the sink, and the spiral source are labeled by black arrows, and the yellow star inside the limit cycle is located at the spiral source. The blue arrows in (b) and the orange arrows in (d) depict the shift direction of the three fixed points as $V$ increases. In (d), vector fields are added in small blue arrows to help show the fixed points. The initial conditions of the solutions are set discretely along the edge, as well as around the spiral source. In (b)-(d), the evolution time $t$ was 0.05. \label{fig:overall}
  }
\end{figure*}

We systematically analyzed the behavior of the circuit by studying its fixed points and limit cycles. For this purpose, we used vector field diagrams of solutions and Jacobi matrices (for more information, see SI Appendix A). This analysis has resulted in the phase diagram
presented in  Fig.~\ref{fig:overall}(a). The diagram  indicates the presence of a global limit cycle in a wide range of the applied voltage, from $V_1'$ to $V_2$, which  accounts for the presence of the spiking regime.
%shows the fixed points and attractors varies around the emergence of the limit cycle.

The main features of the phase diagram are as follows. First of all,
since the behavior is symmetric with respect to the sign of  $V$, we show only the positive voltage region of the phase diagram. Around $V=0$, the only global attractor is a sink. As $V$ increases, at $V_0 \approx 3.4156 $  there is a bifurcation that nucleates a saddle and spiral source; they do not  influence the global attractor, however, until they separate sufficiently at $V_1$ where the saddle splits the phase space  into two disconnected regions, one of which hosts  a limit cycle and the other the original sink that corresponds to a static state, as shown in Fig.~\ref{fig:overall}(b). In SI Appendix B we show the possibility of switching between these two attractors using voltage pulses. Another bifurcation occurs when the sink and saddle point annihilate at $V_1'=7.9582$, transforming the limit cycle into the global attractor with stable spikes, as shown in Fig.~\ref{fig:overall}(c).  At $V_2$, another bifurcation generating a sink-saddle pair cuts off the limit cycle, shifting the global attractor to the sink, as shown in Fig.~\ref{fig:overall}(d). As $V$ continuously increases, the saddle and spiral source point move towards each other and then annihilate, leaving the sink alone (for more information, see SI Appendix A).

\begin{figure}[tb]
  \centering
(a)  \includegraphics[width=7.5cm]{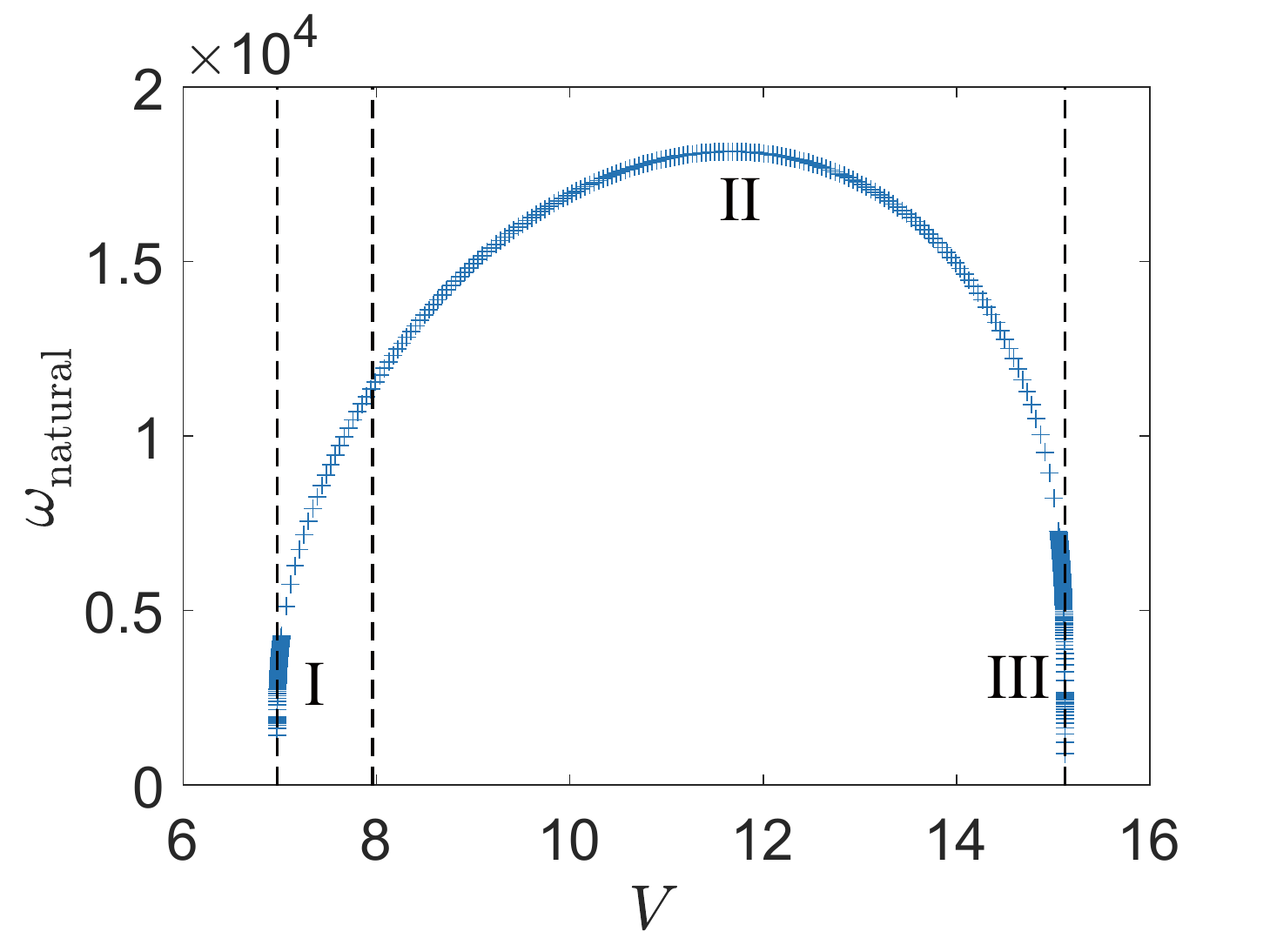} \;\; \;\;\;\;
(b) \includegraphics[width=7.5cm]{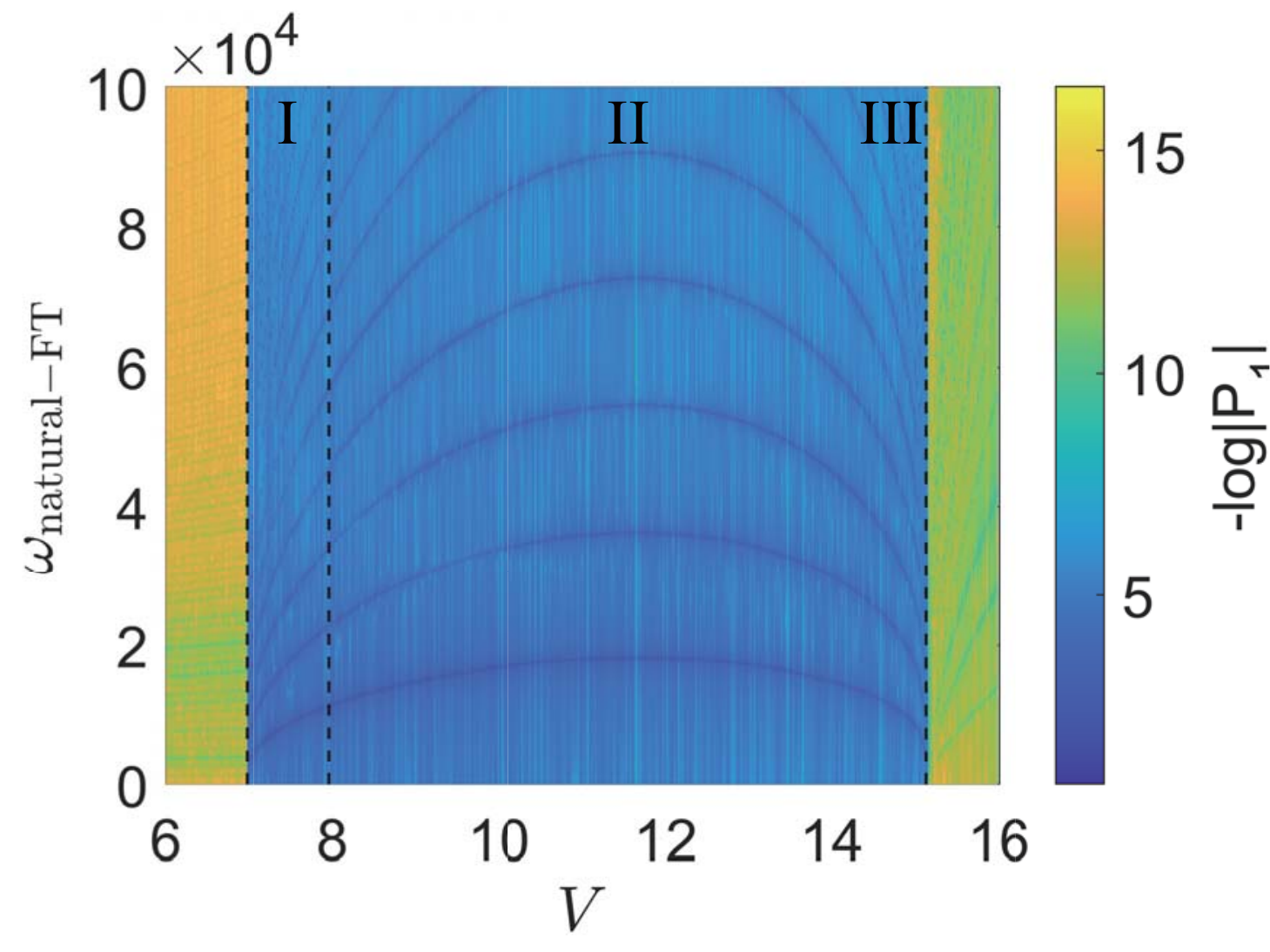} \\
(c)  \includegraphics[width=5cm]{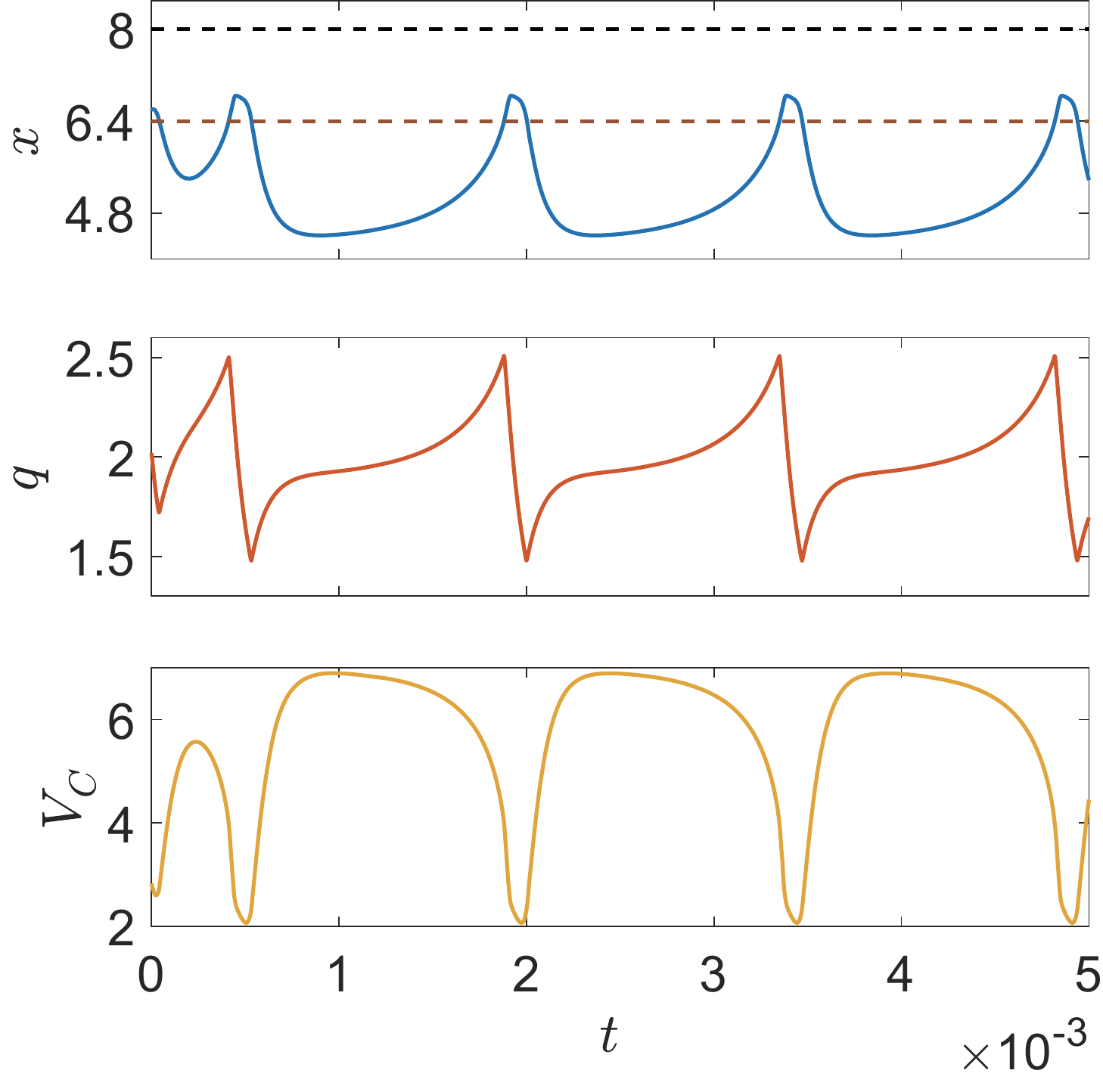}
(d)  \includegraphics[width=5cm]{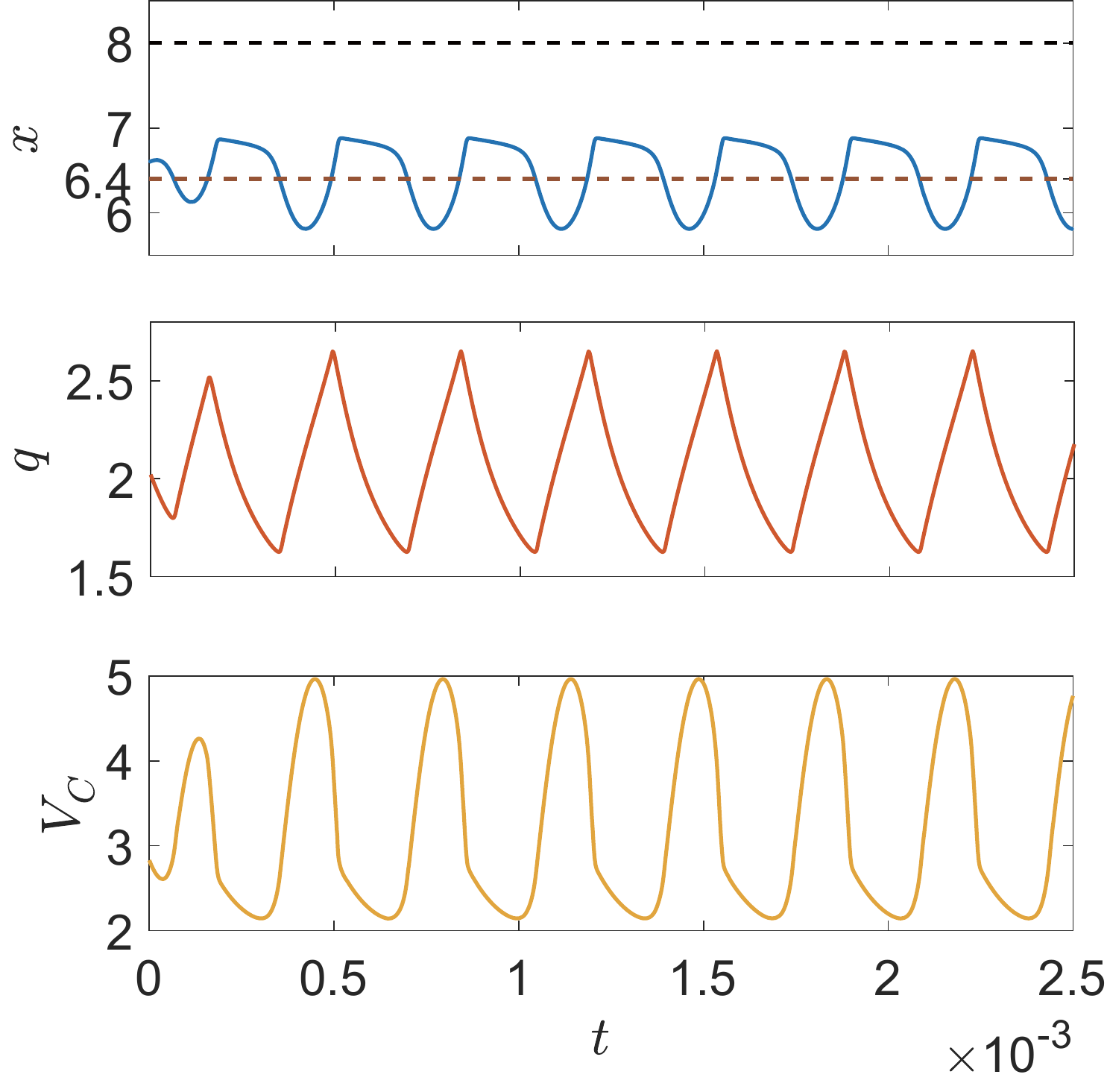}
(e) \includegraphics[width=5cm]{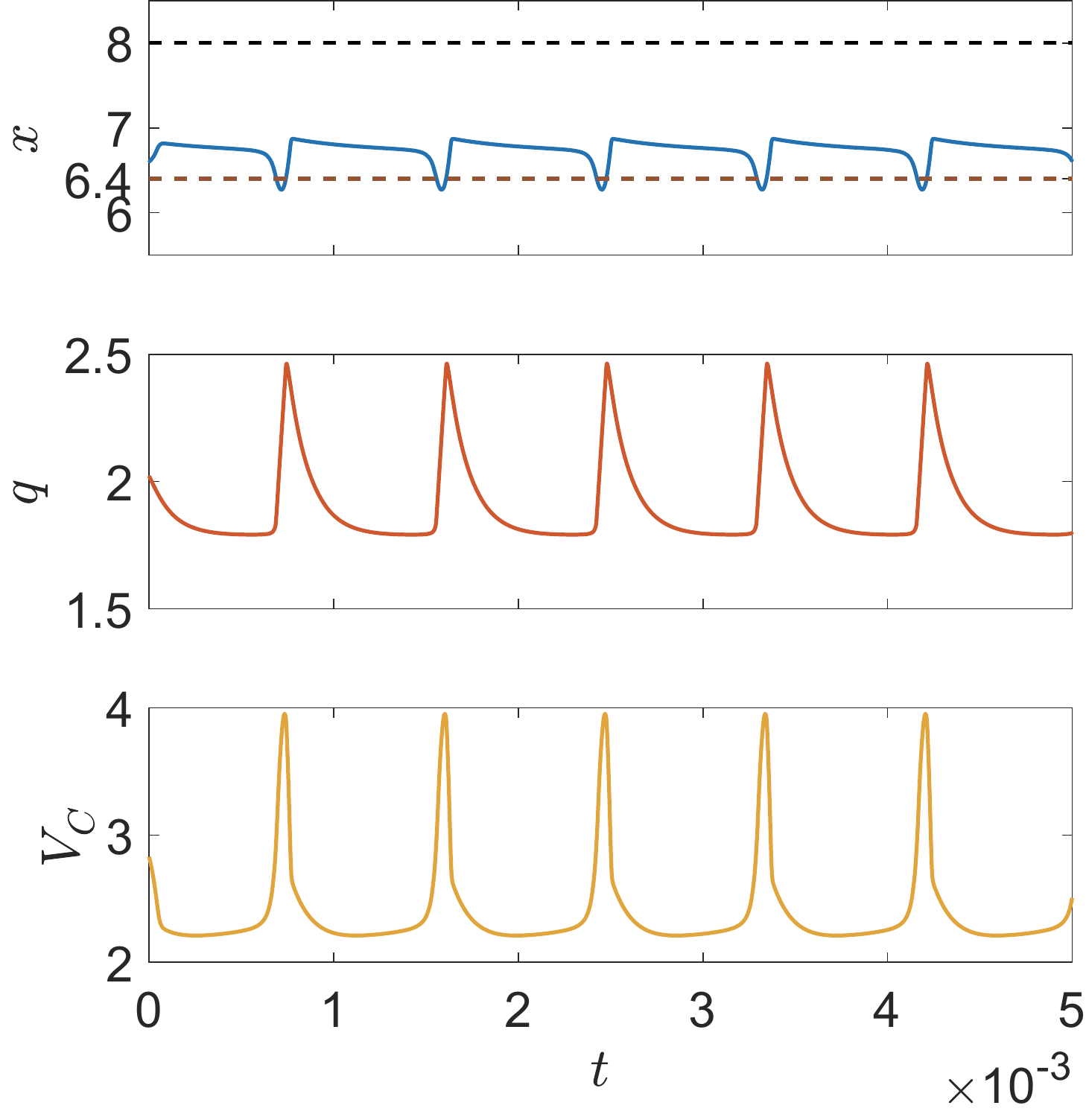}
  \caption{(a) Spike frequency as a function of $V$, and (b) Fourier transform of $V_C$ as a function of $V$. The dashed lines, from left to right, refer to $V_1$, $V_1'$ and $V_2$, respectively. Here, $P_1$ is the single-sided amplitude of the Fourier transform. In these calculations, to keep the system close to the limit cycle, the initial condition was selected as $x_0=6.6000$ and $q_0=2.0207$.
  The evolution time $t$ was 1.5, and we skipped the initial transient interval.
  (c)-(e) Steady-state oscillations at (c) $V=7.0183$ (regime I), (d) $V=11.5470$ (regime II) and (e) $V=15.0561$ (regime III). In (c)-(e), the black dashed line refers to $d$ and the brown one refers to $x_c$.
  }
  \label{natural}
\end{figure}

To better understand the properties of spikes, the \textit{natural} frequency of spikes, $w_{natural}$, was calculated as a function of the applied voltage (see Fig.~\ref{natural}(a)). Interestingly, the calculated points are distributed in the half-of-the-oval shape in the frequency-voltage plot. Fig.~\ref{natural}(a) shows that the frequency approaches zero when $V\rightarrow V_1, V_2$. The Fourier transform of the voltage across the memcapacitor is presented in Fig. \ref{natural}(b). Qualitatively, the whole spiking regime can be divided into three parts, I, II, and III, that are different by the pattern of spikes (see Figs. \ref{natural}(c)-(e)). %The spiky parts of signal $V_C$ are their lower parts in regime I and their upper parts in regime III respectively, while in regime II they are more harmonic.
The ``negative spikes'' regime, I, and the  ``positive spikes'' regime, III, are connected by the regime of more symmetric (harmonic) spikes, II.

\section{Synchronization with external source}

\begin{figure}[ht]
  \centering
(a)\includegraphics[width=5.3cm]{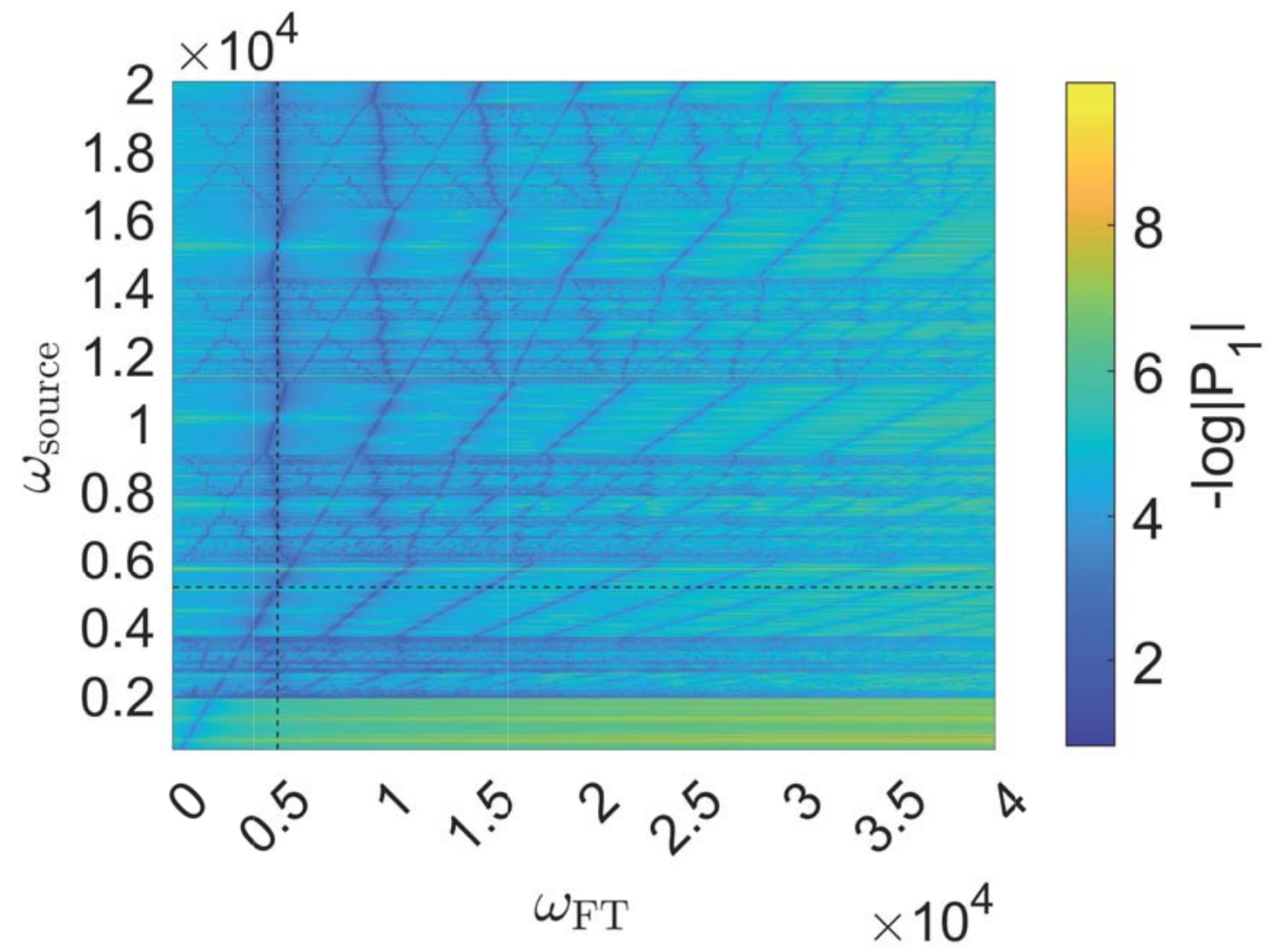}
(b)\includegraphics[width=5.3cm]{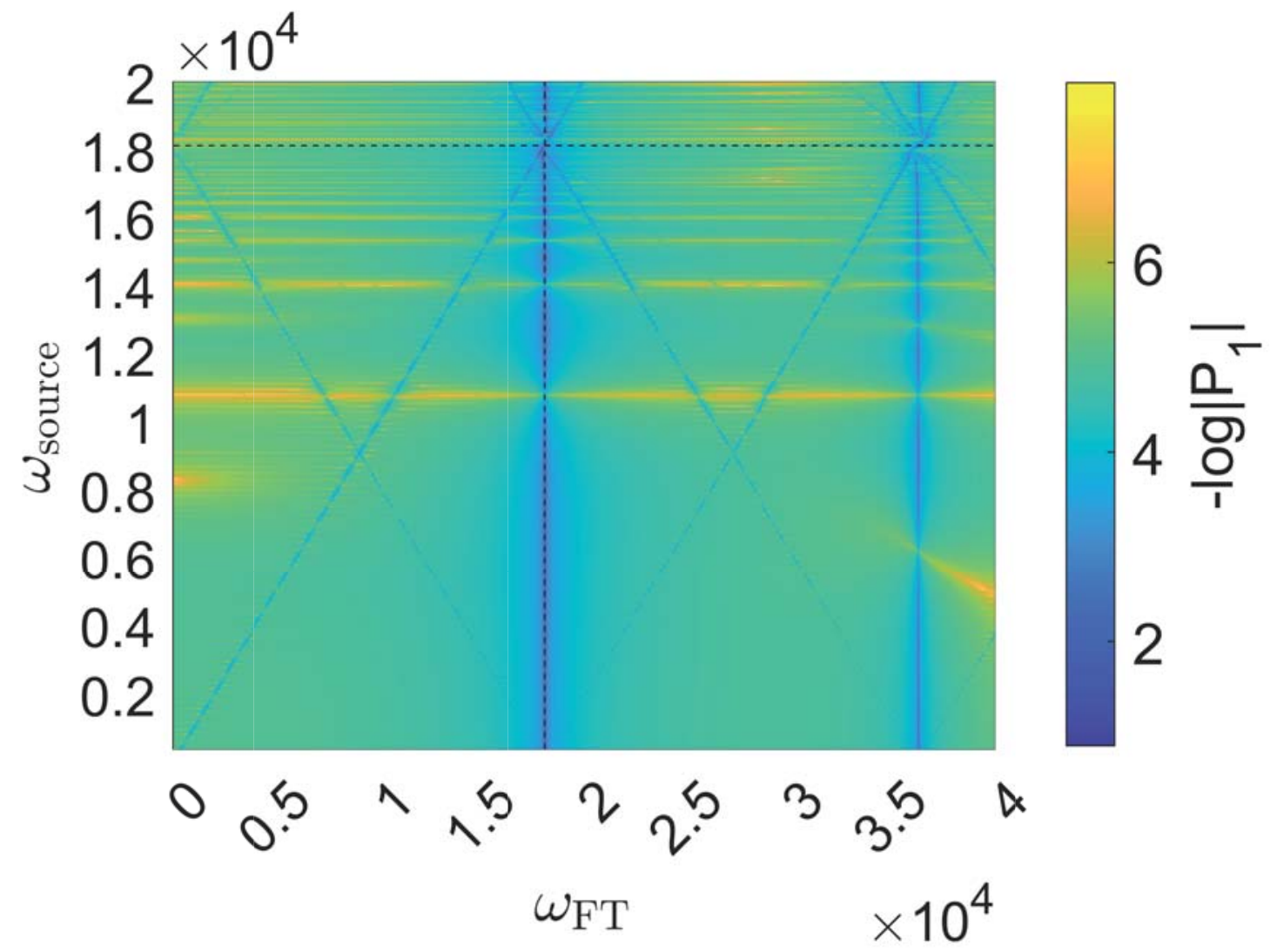}
(c)\includegraphics[width=5.3cm]{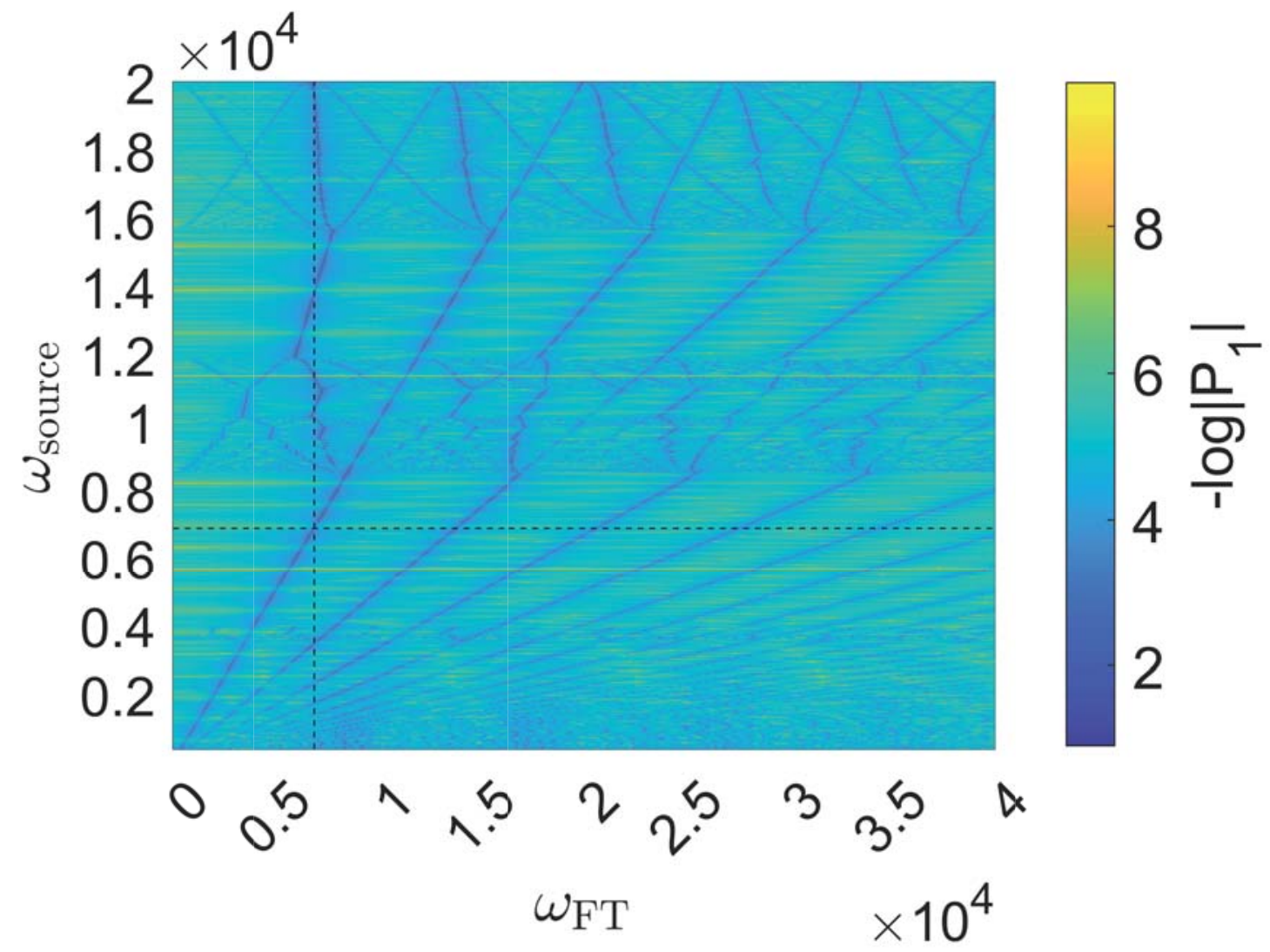}
  \caption{Synchronization with external source: The Fourier transform of $V_C$ at (a) $V_{dc}=7.0645$ (regime I), (b) $V_{dc}=11.5470$ (regime II) and (c) $V_{dc}=15.0688$ (regime III).  $P_1$ is the single-sided amplitude of Fourier transform. In these calculations, to keep the system close to the limit cycle, the initial condition was selected as $x_0=6.6000$ and $q_0=2.0207$. These plots were obtained using the integration time $t$ of 2; the initial interval of transient dynamics was omitted.
  In (a)-(c), the horizontal and vertical dashed lines correspond to $\omega_{natural}(V_{dc})$.
  }
  \label{synFT}
\end{figure}

To study the synchronization with an external source, an ac voltage was added to the constant driving voltage $V_{dc}$, $V(t)=V_{dc}+\delta V\sin(\omega_{source} t)$, with $\delta V=0.1155$. To initialize the system close to the limit cycle, we used the initial conditions $x_0=6.60$ and $q_0=2.02$. The Fourier transforms for the regimes I-III of oscillations in Fig.~\ref{natural}(a) are presented in  Fig. \ref{synFT}. When the circuit is in regime II (Fig. \ref{synFT}(b)), i.e. away from the thresholds, the synchronization occurs only when the source frequency is very close to the spike frequency, $\omega_{source}\approx\omega_{natural}(V_{dc})$.
In this regime, the Fourier plot features various combinations
of integer  multiples of $\omega_{source}$ and $\omega_{natural}$, $N\omega_{natural}\pm M\omega_{source}$, where $M,N=0,1,2,...$

In the other two cases in Fig. \ref{synFT}, the external source has a much stronger influence on the spike generation. Figs. \ref{synFT}(a) and (c) represent the cases when $V_{dc}$ is close to the thresholds $V_1$ and $V_2$.
In these cases, the regions of synchronized driving frequencies are largely extended in both integer multiples and rational fractions of the self-oscillation frequencies. In regime I, when the driving frequency is small, the system responds harmonically to the drive, as shown in Fig. \ref{synFT}(a), where only the component of $\omega_{source}$ is evident. This is because once $V$ drops below $V_1$, the spiral source and the saddle vanish simultaneously, and the system can easily switch to the sink, which is away from the contact regime as depicted in Figs. \ref{fig:overall}(a) and (b), and may not return. In contrast, in regime III, when $V$ crosses over $V_2$, the bifurcation of a sink-saddle pair appears in the contact regime on the previous limit cycle, as depicted in Figs. \ref{fig:overall}(a) and (d), thus the system will not go away from the contact regime, which results in spiking waveform even at low source frequency. Further detailed Fourier transforms and relevant analyses of these three regimes are discussed in SI Appendix C.

\section{Other types of dynamics}

In this section, we show that the dynamics of Fig.~\ref{fig:1}(a) circuit can be further enriched by the use of additional components with memory. Indeed, the replacement of the resistor $r$ in Fig.~\ref{fig:1}(a) with a memristor may completely change the pattern of spikes
 %Below, this approach is illustrated using two memristive models, that both change the regular spiking
 to bursting (a pattern of firing wherein the periods of rapid spiking are separated by quiescent periods).
Below, we introduce two memristive behaviors (without and with threshold with respect to $I_r$) and study their effect on the spike generation.

%We also reproduce some of these adaptations in our model by introducing memory effects.

In principle, it is evident that at a suitable constant applied voltage $V$, the resistance $r$ controls the spike generation. As in the limiting cases of $r\rightarrow 0$ and $r\rightarrow \infty$ the dynamics of Fig.~\ref{fig:1}(a) circuit should not be oscillatory, one can assume that the spiking behavior occurs within certain resistance thresholds, say, when $r_2<r<r_1$. %\textcolor{orange}{$r$ has opposite behavior compared to $V$. To correspond to $V$, should we use $r_2<r<r_1$?}
Consequently, a suitable memristor (e.g., with resistance changing across $r_1$ and/or $r_2$) may be used to control the pattern of spikes.

%Even though several variables in the systems are good candidates for accumulating memory, here we choose the resistor $r$ outside the memcapacitor, since it can be tuned instead of $V$ to control spiking rate.  We mainly focus on driving the system near the spiking thresholds of $r$, namely $r_1$, $r_1'$, and $r_2$, which correspond to $V_1$, $V_1'$, and $V_2$, respectively ($r_1>r_1'>r_2$).

\begin{figure}[ht]
  \centering
  (a)  \includegraphics[width=6.5cm]{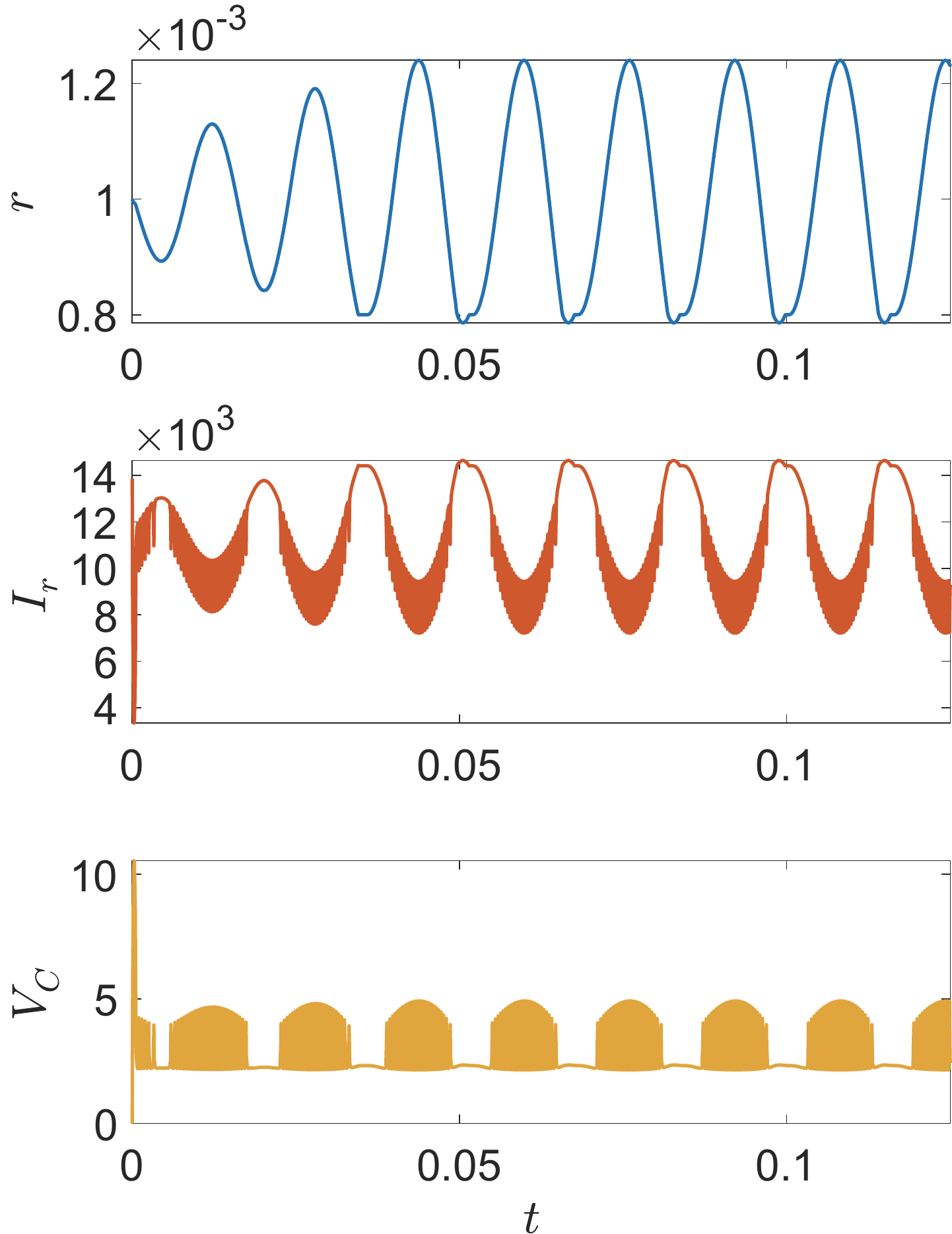} \;\; \;\;\;\;
(b) \includegraphics[width=6.5cm]{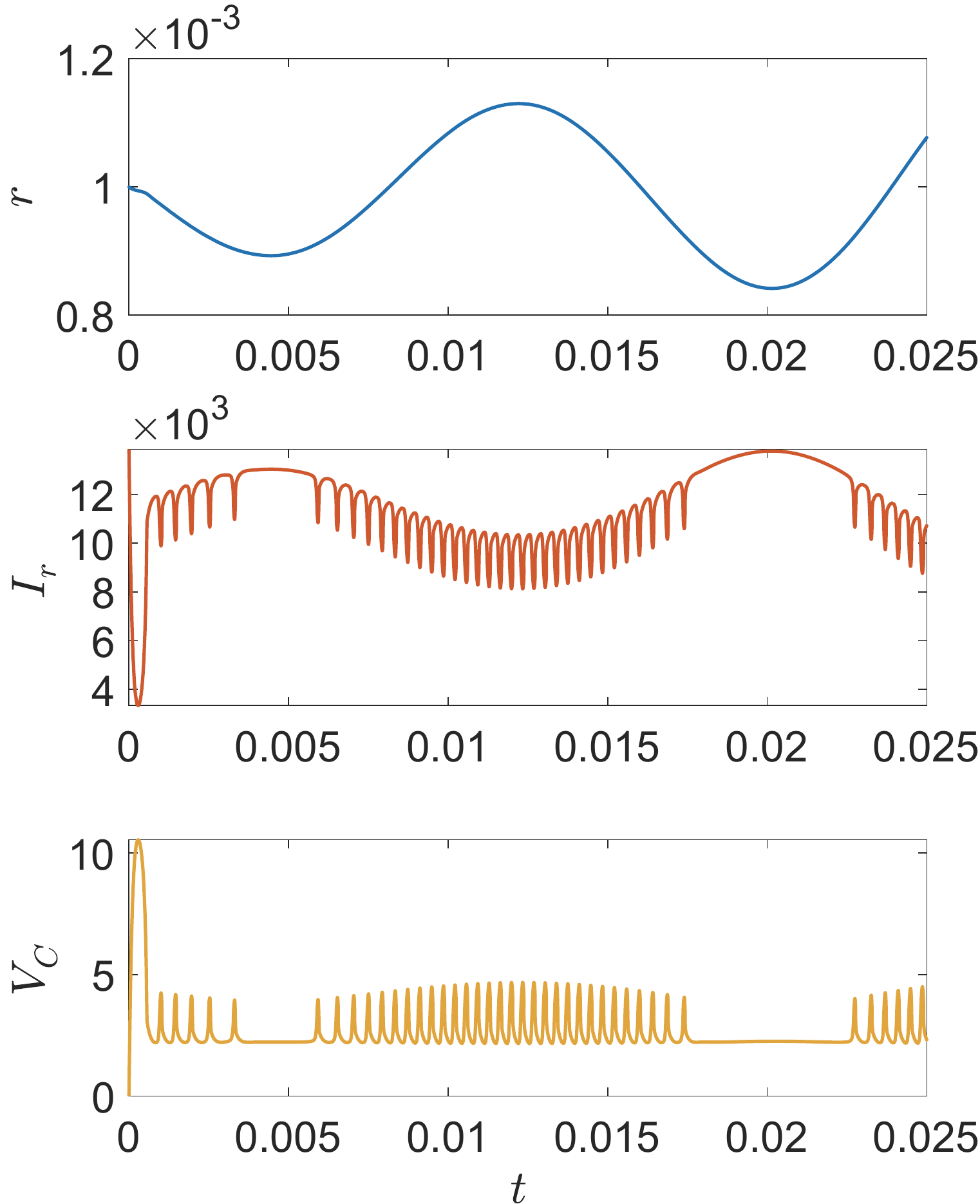} \\
(c)  \includegraphics[width=6.5cm]{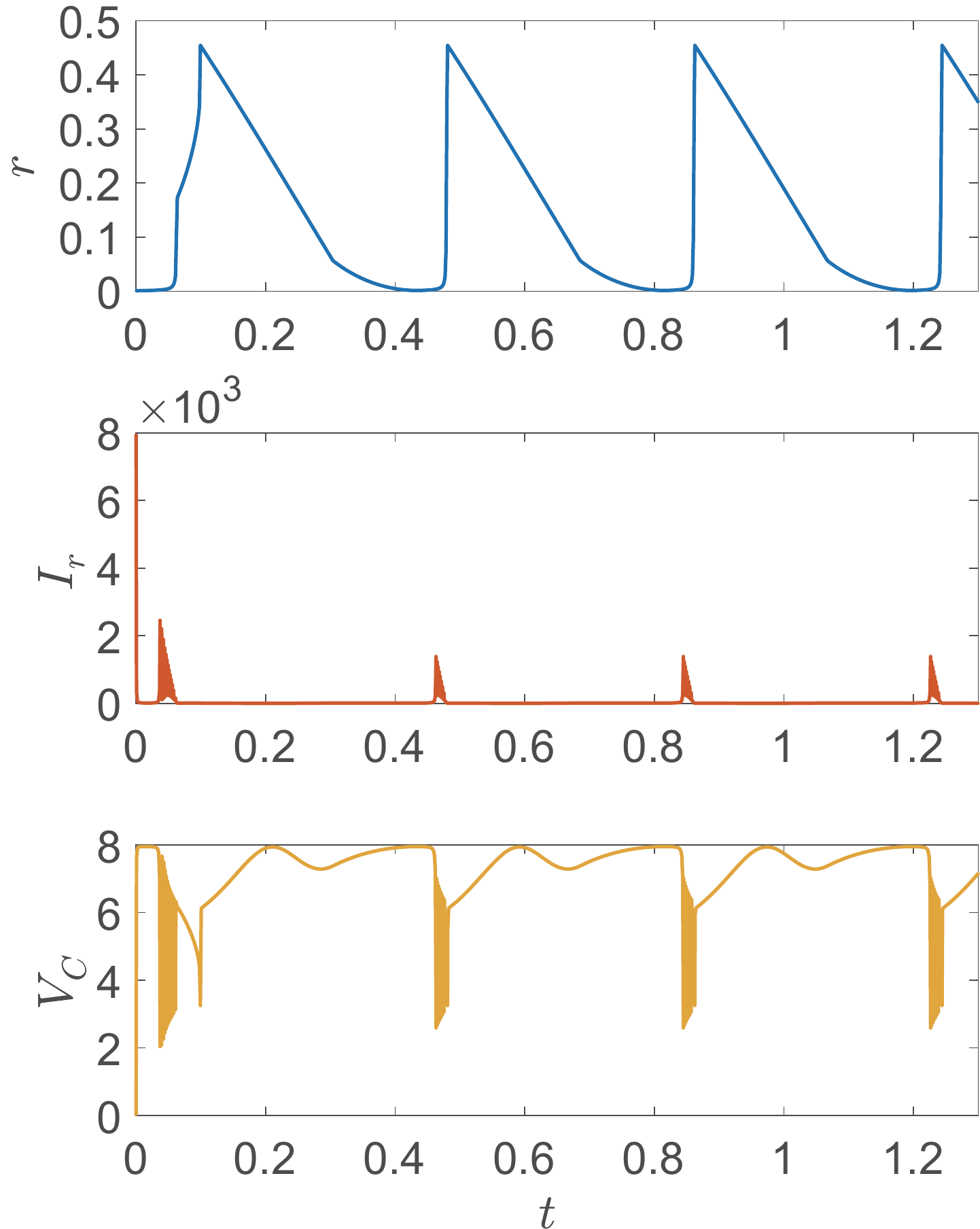} \;\; \;\;\;\;
(d) \includegraphics[width=6.5cm]{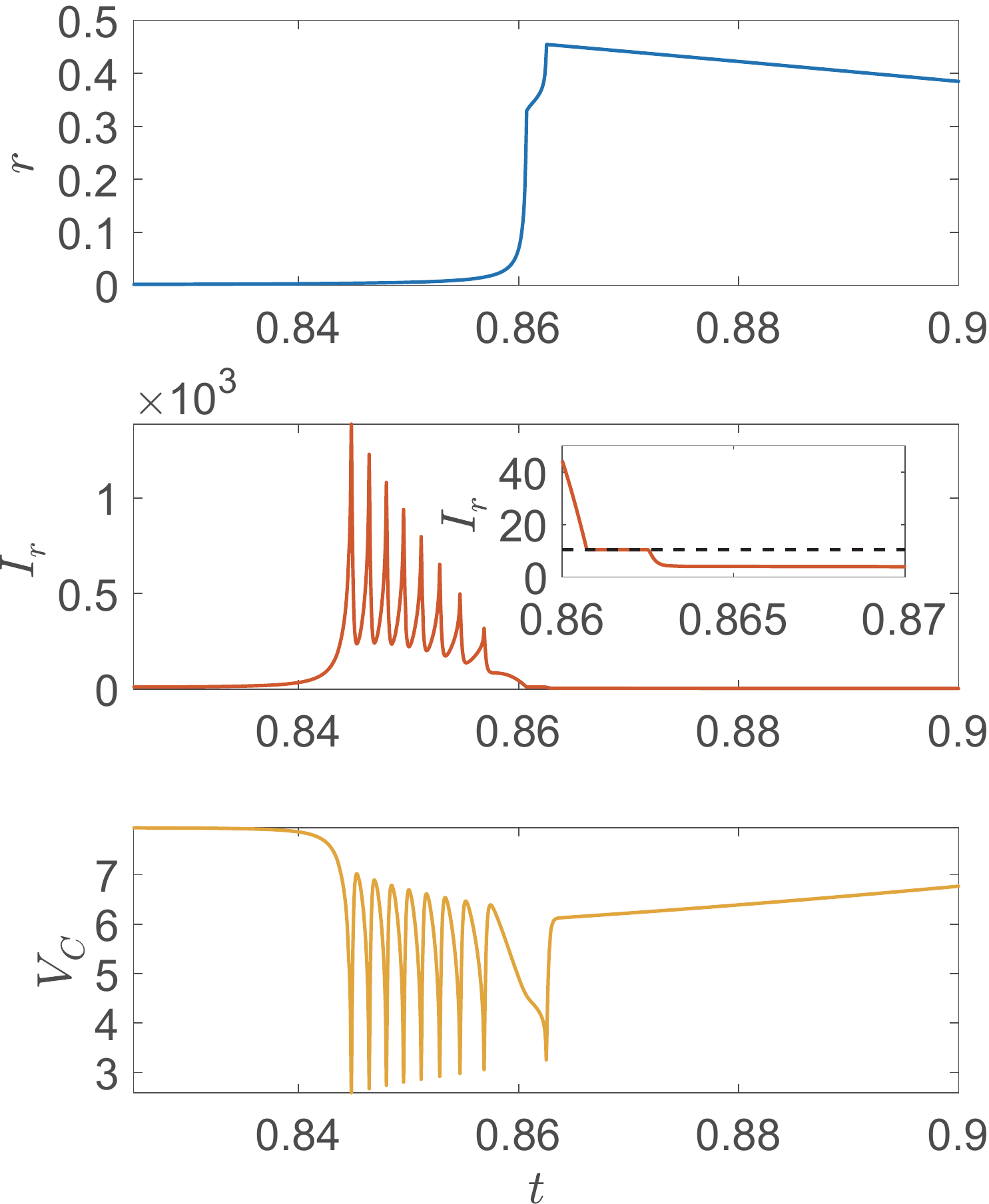}
  \caption{Achieving  complex spiking behaviors by introducing memory into the resistor. (a), (b) Method of Eq. (\ref{eq:adapt1}). (c), (d) Method of Eq. (\ref{eq:adapt2}). Panels (b) and (d) are zoom-ins of  (a) and (c), respectively. In the inset of (d), the dashed line refers to $I'$, and during the short plateau, the limit cycle has just disappeared and the system is switching to the sink. In the simulation (a), the parameters were set as $V=13.8564$, $\alpha_1=3.4641\times10^{-6}$, $\lambda_1=1.600\times10^5$, $r_0=10^{-3}$, $\gamma=0$ and $r'=0.8r_0$. In the simulation (c), $(\text{arctan}\lambda'(I_r(t)-I'))/\pi+0.5$ is used instead of step functions in Eq. (\ref{eq:adapt2}), and the parameters were set as $V=7.9674$, $I'=10.3923$, $\lambda'=1000$, $\alpha_2=1.6000\times10^4$ and $\lambda_2=2000$. Initial conditions are both set $x=0$, $q=0$ and $r_0=10^{-3}$.
  }
  \label{adapt}
\end{figure}

For the sake of simplicity, in what follows the resistance is used as the internal state variable of memristor (such models are known in the literature~\citep{pershin_memristive_2009}). In the first model, the memristor dynamics is defined by the equation
% \begin{equation}
%     \dot{r}(t)=-\alpha_1 I_r(t)+\lambda_1\int\limits_0^te^{-\gamma \left(t-\tau\right)}\left(r_0-r(\tau)\right)\textnormal{d}\tau\;\; ,\label{eq:adapt1}
% \end{equation}
\begin{equation}
    \dot{r}(t)=-\alpha_1 I_r(t)+\lambda_1\int\limits_0^te^{-\gamma \left(t-\tau\right)}\left(r_0-r(\tau)\right)\textnormal{d}\tau\;\; ,\label{eq:adapt1}
\end{equation}
where $\alpha_1$ is the current proportionality coefficient (can be of any sign), $\lambda_1$ is the (positive) relaxation coefficient, $\text{exp}\left(-\gamma \left(t-\tau\right)\right)$ is the memory kernel, $\gamma$ is  non-negative constant,
$r_0$ is the equilibrium resistance, and $I_r(t)=\dot{q}+I_M$ is the current flowing through $r$.
It is assumed that $r$ is confined to the interval from $r'$ to infinity, where $r'$ is the minimal value of $r$.
In the right-hand side of Eq.~(\ref{eq:adapt1}), the first term causes the drift of $r$ mostly due to the spikes. The second term incorporates the relaxation  to $r_0$ and memory on prior states.

% $\theta$ is the Heaviside step function, $r'<r_0$ is the minimum threshold for $r(t)$ to avoid $r(t)\leq0$ if $\alpha_1>0$ \textcolor{orange}{Eq.(8) is modified with the minimum threshold of $r(t)$ here}
%\left(1-\theta\left(r(t)-r'\right)\theta\left(\alpha_1\right)\right)

Figs. \ref{adapt}(a) and (b) show results of simulations with Eq.~(\ref{eq:adapt1}) model for $r$. Roughly the picture is as follows: The spikes starting at $t=0$ cause $r$ to decrease. As soon as $r<r_2$, the spiking terminates and $r$ starts relaxing towards $r_0$ with a delay determined by $\gamma$. When $r$ crosses $r_2$  again (in the opposite direction), the circuit transits back into the spiking mode. This process repeats periodically.

%The first term, $I_r(t)$, is synchronized with spikes, while the second one involves more averaging. The combination of the two helps avoid getting stuck and keeps the system in a cycle between the spiking and non-spiking regimes. Note that between the thresholds $r_1$ and $r_1'$, there are two attractors, so this type of dynamics more easily stabilizes bursting around the other threshold $r_2$. The relevant simulation results  are shown in Fig. \ref{adapt}(a)(b).

Next, we consider the second model for $r(t)$. A deeper analysis of Fig.~\ref{fig:1}(a) circuit (with $r$ as a control parameter) has revealed an additional threshold $r_1'$ between $r_2$ and $r_1$ such that  there are two attractors (a sink and limit cycle) for $r_1>r>r'_1$. %\hl{Correct?} \textcolor{orange}{if using $r_1<r<r_2$, I think it should be $r_2'<r<r_2$. Or if using $r_2<r<r_1$, it should be $r_1'<r<r_1$}.
In the limit cycle, certain variables such as $I_r(t)$ have higher values for the duration of one spiking period compared to those in the sink. Therefore, we can set a threshold, $I'$, to induce the cycling between them. If $I_r(t)>I'$, we can set $\dot{r}>0$,
%\textcolor{red}{I put 1 here, but not sure this is right. But sayng gg 0 is weird} \hl{YP: I think it should be zero as the relation defines the derivative (the direction of change)} \textcolor{orange}{maybe just using $\dot{r}>0$? When $r$ is small, $\dot{r}<0$ may also dominate for some cases},
otherwise $\dot{r}<0$, like
\begin{equation}
    \dot{r}(t)=\left\{
    \begin{aligned}
        \alpha_2 r(t)^2,\quad &I_r(t)>I'\\
        -\lambda_2 r(t), \quad &I_r(t)<I'\\
        0, \quad &I_r(t)=I'
    \end{aligned}
    \right.\;\; ,\label{eq:adapt2}
\end{equation}
where $\alpha_2,\lambda_2$ are positive coefficients. It is assumed that $r$ is confined to the interval from $r'$ to infinity, where $r'$ is the minimal value of $r$. This mechanism allows $r(t)$ to increase from $\{ r_1', r_1\}$ to above $r_1$, switching the system from the limit cycle to the sink; after a while $r(t)$ begins to decrease, closing the cycle. The  simulations  of this scheme for $\dot{r}$ are shown in Fig. \ref{adapt}(c) and (d).
Note that the second method works only when $R_m$ is finite (for more details, see
 SI Appendix A).

\section{Conclusion}

In summary, we have proposed a leaky memcapacitor - an electromechanical crossbreed of memcapacitor and memristor -  that can generate neuromorphic spikes. Its model is based on the potential that combines linear elasticity with nonlinear Lennard-Jones-like interaction between the plates at short distances, attempting to represent realistic interaction potential. Thanks to the presence of the non-linear interaction, the dynamical behavior of the system in the contact region is different from the  one when the plates are relatively far from each other. This helps achieve a stable spiking behavior when a constant voltage is applied.

In order to thoroughly understand the spiking behavior, we have conducted the stability analysis in the $(x,q)$-space and discovered several interesting regimes characterized by different configurations of fixed points and attractors.  We have shown that for some ranges of parameters,  one can use a voltage pulse to switch the system from a sink to a limit cycle and vice-versa. We have also found that the spike shape depends on the applied voltage.

An important feature of the system is that the spike frequency may adapt to the external perturbation frequency (depending on the model and excitation parameters). A rich dynamical behavior has been observed including synchronization when a small amplitude ac signal was added to the constant driving voltage. %Moreover, the synchronization is quite complex near the limit cycle thresholds.
In addition,  replacing constant external resistor by a memristor extends the variety of spike waveforms that the circuit generates.  With this modification,  the circuit can be tuned to mimic  behaviors of some  types of biological neurons.

Overall, the system introduced in this article provides a new avenue for the practical realization of neuromorphic devices based on memcapacitive and memristive effects. Our study may lead to novel  energy-efficient realizations of neural dynamics with electromechanical structures, including artificial analogs of biological membranes.

\section*{Conflict of Interest Statement}
%All financial, commercial or other relationships that might be perceived by the academic community as representing a potential conflict of interest must be disclosed. If no such relationship exists, authors will be asked to confirm the following statement:

The authors declare that the research was conducted in the absence of any commercial or financial relationships that could be construed as a potential conflict of interest.

\section*{Author Contributions}

All authors contributed to conceiving the idea, formulating the model, its qualitative analysis, and writing the manuscript. Z.~Z. has performed all  numerical simulations.

\section*{Funding}
I.~M. acknowledges funding from
the Materials Sciences and Engineering Division, Basic Energy Sciences, Office of Science, US DOE.

%\section*{Acknowledgments}
%This is a short text to acknowledge the contributions of specific colleagues, institutions, or agencies that aided the efforts of the authors.

\section*{Data Availability Statement}
The Supplemental Information contains additional data supporting this manuscript.
Further inquiries can be directed to the corresponding authors.

\bibliographystyle{Frontiers-Harvard} %  Many Frontiers journals use the Harvard referencing system (Author-date), to find the style and resources for the journal you are submitting to: https://zendesk.frontiersin.org/hc/en-us/articles/360017860337-Frontiers-Reference-Styles-by-Journal. For Humanities and Social Sciences articles please include page numbers in the in-text citations
%\bibliographystyle{Frontiers-Vancouver} % Many Frontiers journals use the numbered referencing system, to find the style and resources for the journal you are submitting to: https://zendesk.frontiersin.org/hc/en-us/articles/360017860337-Frontiers-Reference-Styles-by-Journal
%\bibliography{test}
\bibliography{ref}

\begin{thebibliography}{40}
\providecommand{\natexlab}[1]{#1}
\expandafter\ifx\csname urlstyle\endcsname\relax
  \providecommand{\doi}[1]{doi:\discretionary{}{}{}#1}\else
  \providecommand{\doi}{doi:\discretionary{}{}{}\begingroup
  \urlstyle{rm}\Url}\fi
\providecommand{\selectlanguage}[1]{\relax}
\providecommand{\bibAnnoteFile}[1]{%
  \IfFileExists{#1}{\begin{quotation}\noindent\textsc{Key:} #1\\
  \textsc{Annotation:}\ \input{#1}\end{quotation}}{}}
\providecommand{\bibAnnote}[2]{%
  \begin{quotation}\noindent\textsc{Key:} #1\\
  \textsc{Annotation:}\ #2\end{quotation}}

\bibitem[{Abbott(1999)}]{abbott_lapicques_1999}
Abbott, L.~F. (1999).
\newblock Lapicque’s introduction of the integrate-and-fire model neuron
  (1907).
\newblock \emph{Brain Research Bulletin} 50, 303--304
\bibAnnoteFile{abbott_lapicques_1999}

\bibitem[{Andersen et~al.(2009)Andersen, Jackson, and
  Heimburg}]{andersen_towards_2009}
Andersen, S. S.~L., Jackson, A.~D., and Heimburg, T. (2009).
\newblock Towards a thermodynamic theory of nerve pulse propagation.
\newblock \emph{Progress in Neurobiology} 88, 104--113
\bibAnnoteFile{andersen_towards_2009}

\bibitem[{Appali et~al.(2012)Appali, van Rienen, and
  Heimburg}]{appali_comparison_2012}
Appali, R., van Rienen, U., and Heimburg, T. (2012).
\newblock A comparison of the {Hodgkin}–{Huxley} model and the soliton theory
  for the action potential in nerves.
\newblock In \emph{Advances in {Planar} {Lipid} {Bilayers} and {Liposomes}}
  (Elsevier), vol.~16. 275--299
\bibAnnoteFile{appali_comparison_2012}

\bibitem[{Bryant and Segundo(1976)}]{bryant_spike_1976}
Bryant, H.~L. and Segundo, J.~P. (1976).
\newblock Spike initiation by transmembrane current: a white-noise analysis.
\newblock \emph{The Journal of Physiology} 260, 279--314
\bibAnnoteFile{bryant_spike_1976}

\bibitem[{Catterall et~al.(2012)Catterall, Raman, Robinson, Sejnowski, and
  Paulsen}]{catterall_hodgkin-huxley_2012}
Catterall, W.~A., Raman, I.~M., Robinson, H. P.~C., Sejnowski, T.~J., and
  Paulsen, O. (2012).
\newblock The {Hodgkin}-{Huxley} heritage: {From} channels to circuits.
\newblock \emph{Journal of Neuroscience} 32, 14064--14073
\bibAnnoteFile{catterall_hodgkin-huxley_2012}

\bibitem[{Chen et~al.(2019)Chen, Garcia-Gonzalez, and
  Jérusalem}]{chen_computational_2019}
Chen, H., Garcia-Gonzalez, D., and Jérusalem, A. (2019).
\newblock Computational model of the mechanoelectrophysiological coupling in
  axons with application to neuromodulation.
\newblock \emph{Phys. Rev. E} 99, 032406
\bibAnnoteFile{chen_computational_2019}

\bibitem[{Chua et~al.(2012)Chua, Sbitnec, and Kim}]{chua_hodgkinhuxley_2012}
Chua, L., Sbitnec, V., and Kim, H. (2012).
\newblock {H}odgkin–{H}uxley axon is made of memristors.
\newblock \emph{International Journal of Bifurcation and Chaos} 22, 1230011
\bibAnnoteFile{chua_hodgkinhuxley_2012}

\bibitem[{Chua and Kang(1976)}]{chua76a}
Chua, L.~O. and Kang, S.~M. (1976).
\newblock Memristive devices and systems.
\newblock \emph{Proceedings of {IEEE}} 64, 209--223
\bibAnnoteFile{chua76a}

\bibitem[{Demasius et~al.(2021)Demasius, Kirschen, and Parkin}]{parkin21a}
Demasius, K.-U., Kirschen, A., and Parkin, S. (2021).
\newblock Energy-efficient memcapacitor devices for neuromorphic computing.
\newblock \emph{Nature Electronics} 4, 748--756
\bibAnnoteFile{parkin21a}

\bibitem[{Di~Ventra and Pershin(2013)}]{diventra13a}
Di~Ventra, M. and Pershin, Y.~V. (2013).
\newblock The parallel approach.
\newblock \emph{Nature Physics} 9, 200
\bibAnnoteFile{diventra13a}

\bibitem[{Di~Ventra et~al.(2009)Di~Ventra, Pershin, and
  Chua}]{di_ventra_circuit_2009}
Di~Ventra, M., Pershin, Y.~V., and Chua, L.~O. (2009).
\newblock Circuit {Elements} {With} {Memory}: {Memristors}, {Memcapacitors},
  and {Meminductors}.
\newblock \emph{Proceedings of the IEEE} 97, 1717--1724
\bibAnnoteFile{di_ventra_circuit_2009}

\bibitem[{Fuortes and Mantegazzini(1962)}]{fuortes_interpretation_1962}
Fuortes, M. G.~F. and Mantegazzini, F. (1962).
\newblock Interpretation of the repetitive firing of nerve cells.
\newblock \emph{Journal of General Physiology} 45, 1163--1179.
\newblock \doi{10.1085/jgp.45.6.1163}
\bibAnnoteFile{fuortes_interpretation_1962}

\bibitem[{Galassi and Wilke(2021)}]{galassi_coupling_2021}
Galassi, V.~V. and Wilke, N. (2021).
\newblock On the {Coupling} between {Mechanical} {Properties} and
  {Electrostatics} in {Biological} {Membranes}.
\newblock \emph{Membranes} 11, 478
\bibAnnoteFile{galassi_coupling_2021}

\bibitem[{Gerstner et~al.(2014)Gerstner, Kistler, Naud, and
  Paninski}]{gerstner_neuronal_2014}
Gerstner, W., Kistler, W.~M., Naud, R., and Paninski, L. (2014).
\newblock \emph{Neuronal Dynamics: {From} Single Neurons to Networks and Models
  of Cognition} (Cambridge University Press)
\bibAnnoteFile{gerstner_neuronal_2014}

\bibitem[{Heimburg(2012)}]{heimburg_capacitance_2012}
Heimburg, T. (2012).
\newblock The {Capacitance} and {Electromechanical} {Coupling} of {Lipid}
  {Membranes} {Close} to {Transitions}: {The} {Effect} of {Electrostriction}.
\newblock \emph{Biophysical Journal} 103, 918--929
\bibAnnoteFile{heimburg_capacitance_2012}

\bibitem[{Heimburg and Jackson(2007)}]{heimburg_thermodynamics_2007}
Heimburg, T. and Jackson, A.~D. (2007).
\newblock The thermodynamics of general anesthesia.
\newblock \emph{Biophysical journal} 92, 3159--3165
\bibAnnoteFile{heimburg_thermodynamics_2007}

\bibitem[{Hodgkin and Huxley(1952{\natexlab{a}})}]{hodgkin_components_1952}
Hodgkin, A.~L. and Huxley, A.~F. (1952{\natexlab{a}}).
\newblock The components of membrane conductance in the giant axon of {Loligo}.
\newblock \emph{The Journal of Physiology} 116, 473--496
\bibAnnoteFile{hodgkin_components_1952}

\bibitem[{Hodgkin and Huxley(1952{\natexlab{b}})}]{hodgkin_currents_1952}
Hodgkin, A.~L. and Huxley, A.~F. (1952{\natexlab{b}}).
\newblock Currents carried by sodium and potassium ions through the membrane of
  the giant axon of {Loligo}.
\newblock \emph{The Journal of Physiology} 116, 449--472
\bibAnnoteFile{hodgkin_currents_1952}

\bibitem[{Hodgkin and Huxley(1952{\natexlab{c}})}]{hodgkin_dual_1952}
Hodgkin, A.~L. and Huxley, A.~F. (1952{\natexlab{c}}).
\newblock The dual effect of membrane potential on sodium conductance in the
  giant axon of {Loligo}.
\newblock \emph{The Journal of Physiology} 116, 497--506
\bibAnnoteFile{hodgkin_dual_1952}

\bibitem[{Hodgkin and Huxley(1952{\natexlab{d}})}]{hodgkin_quantitative_1952}
Hodgkin, A.~L. and Huxley, A.~F. (1952{\natexlab{d}}).
\newblock A quantitative description of membrane current and its application to
  conduction and excitation in nerve.
\newblock \emph{The Journal of Physiology} 117, 500--544
\bibAnnoteFile{hodgkin_quantitative_1952}

\bibitem[{Hodgkin et~al.(1952)Hodgkin, Huxley, and
  Katz}]{hodgkin_measurement_1952}
Hodgkin, A.~L., Huxley, A.~F., and Katz, B. (1952).
\newblock Measurement of current-voltage relations in the membrane of the giant
  axon of {Loligo}.
\newblock \emph{The Journal of Physiology} 116, 424--448
\bibAnnoteFile{hodgkin_measurement_1952}

\bibitem[{Holland et~al.(2019)Holland, de~Regt, and
  Drukarch}]{holland_thinking_2019}
Holland, L., de~Regt, H.~W., and Drukarch, B. (2019).
\newblock Thinking about the nerve {Impulse}: {The} prospects for the
  development of a comprehensive account of nerve impulse {Propagation}.
\newblock \emph{Frontiers in Cellular Neuroscience} 13, 208
\bibAnnoteFile{holland_thinking_2019}

\bibitem[{Izhikevich(2003)}]{izhikevich_simple_2003}
Izhikevich, E. (2003).
\newblock Simple model of spiking neurons.
\newblock \emph{IEEE Transactions on Neural Networks} 14, 1569--1572
\bibAnnoteFile{izhikevich_simple_2003}

\bibitem[{Izhikevich(2007)}]{izhikevich_dynamical_2007}
Izhikevich, E.~M. (2007).
\newblock \emph{Dynamical systems in neuroscience: the geometry of excitability
  and bursting}.
\newblock Computational neuroscience (Cambridge, Mass: MIT Press)
\bibAnnoteFile{izhikevich_dynamical_2007}

\bibitem[{Jing and Das(2018)}]{jing_electric_2018}
Jing, H. and Das, S. (2018).
\newblock Electric double layer electrostatics of lipid-bilayer-encapsulated
  nanoparticles: {Toward} a better understanding of protocell electrostatics.
\newblock \emph{Electrophoresis} 39, 752--759
\bibAnnoteFile{jing_electric_2018}

\bibitem[{Johnson and Winlow(2018)}]{johnson_soliton_2018}
Johnson, A.~S. and Winlow, W. (2018).
\newblock The soliton and the action potential – primary elements underlying
  sentience.
\newblock \emph{Frontiers in Physiology} 9
\bibAnnoteFile{johnson_soliton_2018}

\bibitem[{Kang et~al.(2016)Kang, Huang, and Zhou}]{kang_dynamic_2016}
Kang, Q., Huang, B., and Zhou, M. (2016).
\newblock Dynamic behavior of artificial {Hodgkin}–{Huxley} neuron model
  subject to additive noise.
\newblock \emph{IEEE Transactions on Cybernetics} 46, 2083--2093
\bibAnnoteFile{kang_dynamic_2016}

\bibitem[{Koch and Segev(1998)}]{koch_methods_1998}
Koch, C. and Segev, I. (1998).
\newblock \emph{Methods in neuronal modeling: from ions to networks} (MIT
  press)
\bibAnnoteFile{koch_methods_1998}

\bibitem[{Ling et~al.(2020)Ling, Boyle, Zuckerman, Flores, Ramakrishnan,
  Deisseroth et~al.}]{ling2020high}
Ling, T., Boyle, K.~C., Zuckerman, V., Flores, T., Ramakrishnan, C.,
  Deisseroth, K., et~al. (2020).
\newblock High-speed interferometric imaging reveals dynamics of neuronal
  deformation during the action potential.
\newblock \emph{Proceedings of the National Academy of Sciences} 117,
  10278--10285
\bibAnnoteFile{ling2020high}

\bibitem[{Liu et~al.(2020)Liu, Dong, Qin, and Yan}]{liu_new_2020}
Liu, R., Dong, R., Qin, S., and Yan, X. (2020).
\newblock A new type artificial synapse based on the organic copolymer
  memcapacitor.
\newblock \emph{Organic Electronics} 81, 105680
\bibAnnoteFile{liu_new_2020}

\bibitem[{Martinez-Rincon et~al.(2010)Martinez-Rincon, Di~Ventra, and
  Pershin}]{martinez2010solid}
Martinez-Rincon, J., Di~Ventra, M., and Pershin, Y.~V. (2010).
\newblock Solid-state memcapacitive system with negative and diverging
  capacitance.
\newblock \emph{Physical Review B} 81, 195430
\bibAnnoteFile{martinez2010solid}

\bibitem[{Martinez-Rincon and Pershin(2011)}]{martinez2011bistable}
Martinez-Rincon, J. and Pershin, Y.~V. (2011).
\newblock Bistable nonvolatile elastic-membrane memcapacitor exhibiting a
  chaotic behavior.
\newblock \emph{IEEE {T}ransactions on {E}lectron {D}evices} 58, 1809--1812
\bibAnnoteFile{martinez2011bistable}

\bibitem[{Najem et~al.(2019)Najem, Hasan, Williams, Weiss, Rose, Taylor
  et~al.}]{najem_dynamical_2019}
Najem, J.~S., Hasan, M.~S., Williams, R.~S., Weiss, R.~J., Rose, G.~S., Taylor,
  G.~J., et~al. (2019).
\newblock Dynamical nonlinear memory capacitance in biomimetic membranes.
\newblock \emph{Nature Communications} 10, 3239
\bibAnnoteFile{najem_dynamical_2019}

\bibitem[{Pershin and {Di Ventra}(2010)}]{pershin09c}
Pershin, Y.~V. and {Di Ventra}, M. (2010).
\newblock Experimental demonstration of associative memory with memristive
  neural networks.
\newblock \emph{Neural {N}etworks} 23, 881
\bibAnnoteFile{pershin09c}

\bibitem[{Pershin and Di~Ventra(2011)}]{pershin_memory_2011}
Pershin, Y.~V. and Di~Ventra, M. (2011).
\newblock Memory effects in complex materials and nanoscale systems.
\newblock \emph{Advances in Physics} 60, 145--227
\bibAnnoteFile{pershin_memory_2011}

\bibitem[{Pershin and Di~Ventra(2014)}]{pershin2014memcapacitive}
Pershin, Y.~V. and Di~Ventra, M. (2014).
\newblock Memcapacitive neural networks.
\newblock \emph{Electronics letters} 50, 141--143
\bibAnnoteFile{pershin2014memcapacitive}

\bibitem[{Pershin et~al.(2009)Pershin, La~Fontaine, and
  Di~Ventra}]{pershin_memristive_2009}
Pershin, Y.~V., La~Fontaine, S., and Di~Ventra, M. (2009).
\newblock Memristive model of amoeba learning.
\newblock \emph{Phys. Rev. E} 80, 021926
\bibAnnoteFile{pershin_memristive_2009}

\bibitem[{Scott et~al.(2022)Scott, Bolmatov, Podar, Liu, Kinnun, Doughty
  et~al.}]{scott2022evidence}
Scott, H.~L., Bolmatov, D., Podar, P.~T., Liu, Z., Kinnun, J.~J., Doughty, B.,
  et~al. (2022).
\newblock Evidence for long-term potentiation in phospholipid membranes.
\newblock \emph{Proceedings of the National Academy of Sciences} 119,
  e2212195119
\bibAnnoteFile{scott2022evidence}

\bibitem[{Strogatz(2018)}]{strogatz2018nonlinear}
Strogatz, S.~H. (2018).
\newblock \emph{Nonlinear dynamics and chaos: with applications to physics,
  biology, chemistry, and engineering} (CRC press)
\bibAnnoteFile{strogatz2018nonlinear}

\bibitem[{Zhou and Kurths(2003)}]{zhou_noise-induced_2003}
Zhou, C. and Kurths, J. (2003).
\newblock Noise-induced synchronization and coherence resonance of a
  {Hodgkin}–{Huxley} model of thermally sensitive neurons.
\newblock \emph{Chaos: An Interdisciplinary Journal of Nonlinear Science} 13,
  401--409
\bibAnnoteFile{zhou_noise-induced_2003}

\end{thebibliography}

%%% Make sure to upload the bib file along with the tex file and PDF
%%% Please see the test.bib file for some examples of references
\newpage

\section*{Supplemental Information}

\section*{Appendix A: Phase space analysis}

\begin{figure}[h]
  \centering
  \includegraphics[width=8.0cm]{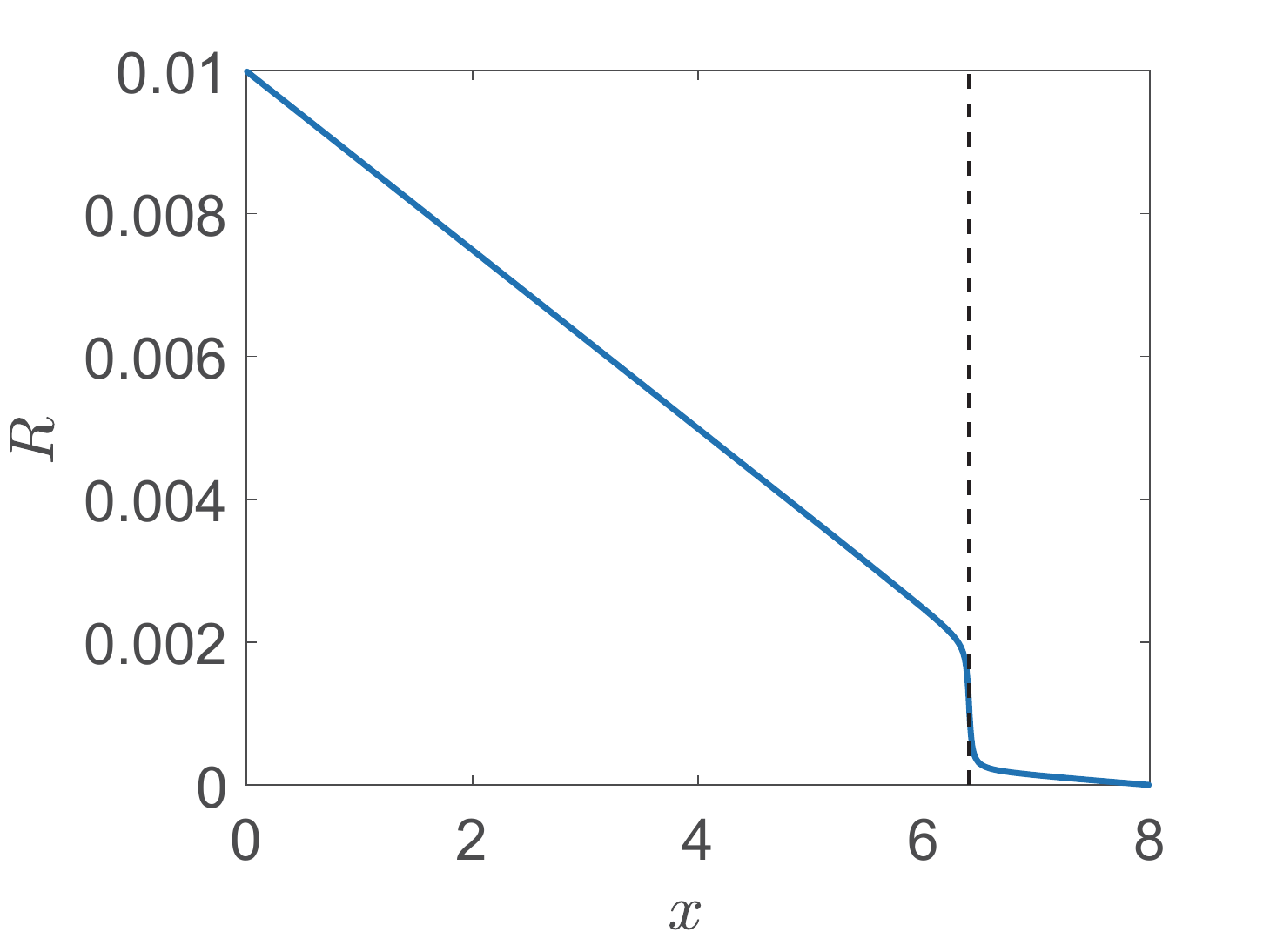}
  \caption{Resistance as a function of the displacement of the top membrane in type II model, Eq.~(\ref{eq:RtypeII}). The dashed line corresponds to $x_c$.
  }
  \label{RII}
\end{figure}

Here we provide details of the analysis of fixed points and attractors in the leaky memcapacitor circuit driven by constant voltage $V$.
As a part of this analysis, we investigated the sensitivity of our results to the choice of the memristive component in the model, $R(x)$. For the sensitivity analysis,  we have modified the equation for $R(x)$ in the main text (``type I" memristance) to the following one
\begin{equation}
    R(x)=\rho_0\frac{d-x}{A}\cdot\left[\left(\frac{1}{\pi}\text{arctan}\beta\left(x_c-x\right)+0.5\right)\cdot9+1\right]\;\;
    \label{eq:RtypeII}
\end{equation}
(``type II" memristance),  which is represented in Fig.~\ref{RII}. For this type of $R(x)$, we used $\beta=50$ in numerical simulations.

\begin{figure}[h]
  \centering
  \includegraphics[width=17.0cm]{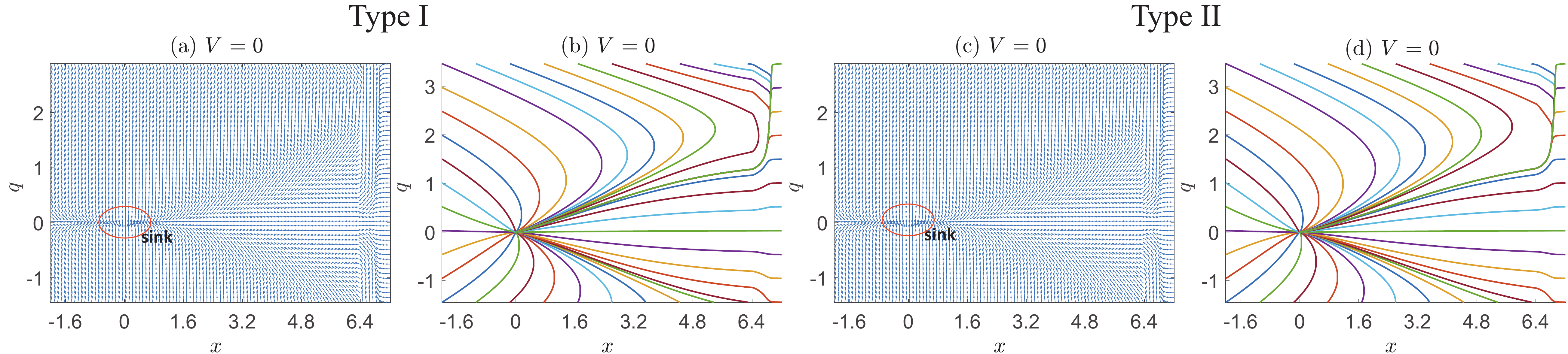}
  \caption{(a), (c) Normalized flow fields and (b), (d) trajectories at $V=0$. To obtain these plots, the initial conditions were set at equal spacing along the edge of the plots. The fixed points are encircled in (a) and (c). Note that in both cases, the only fixed point is a sink.
  }
  \label{th0}
\end{figure}

\begin{figure}[h]
  \centering
  \includegraphics[width=17.0cm]{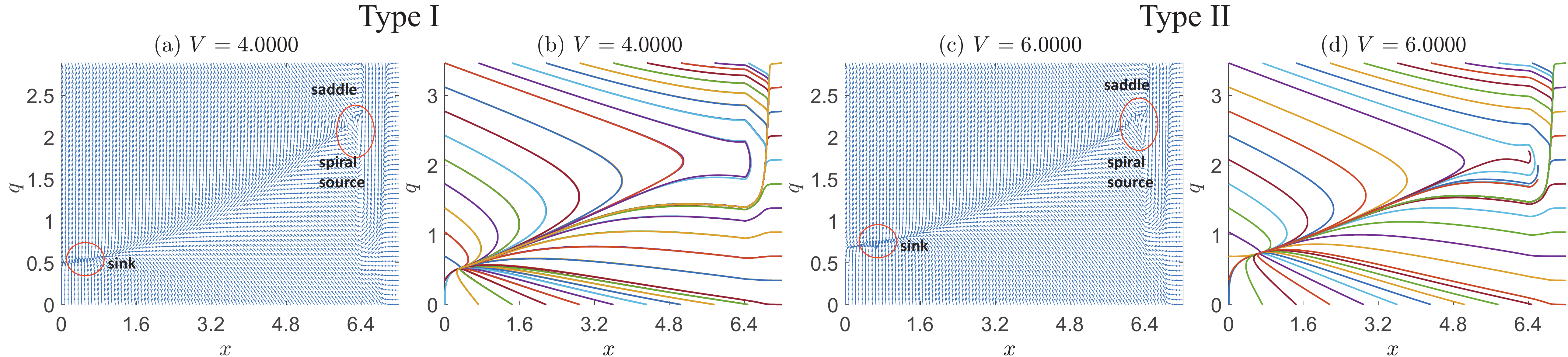}
  \caption{(a), (c) Normalized flow fields and (b), (d) trajectories for $0<V<V_1$. To obtain these plots, the initial conditions were set at equal spacing  along the edge of the plots as well as around spiral sources. The fixed points are encircled in (a) and (c). Note that in both cases, the fixed points are a sink, saddle, and spiral source.
}
  \label{th1}
\end{figure}

\begin{figure}[h]
  \centering
  \includegraphics[width=17.0cm]{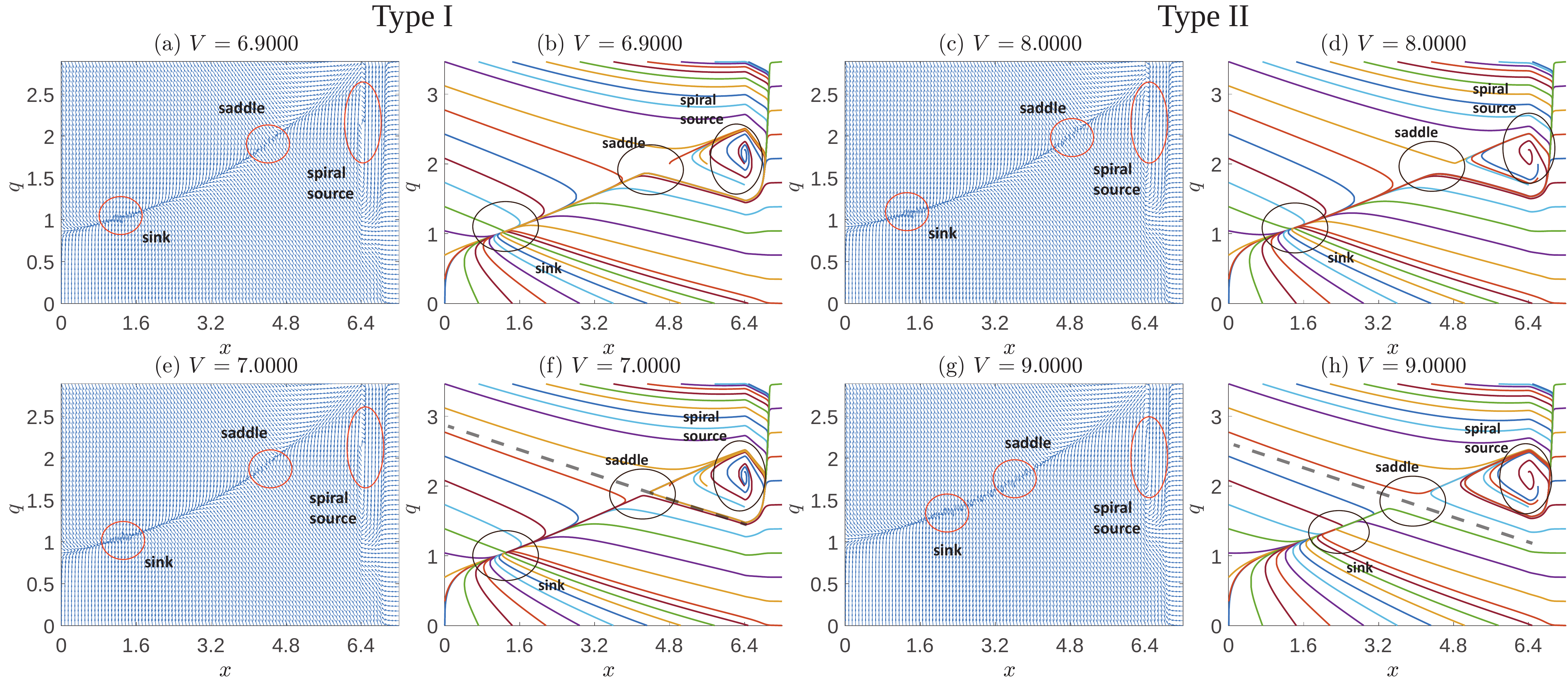}
  \caption{(a), (c), (e), (g) Normalized flow fields and (b), (d), (f), (h) trajectories for two values of $V$ surrounding $V_1$.
  Note that in (f) and (g), the phase space is divided into two parts dominated by a sink and limit cycle, as depicted by dashed lines.
  To obtain these plots, the initial conditions were set at equal spacing  along the edge of the plots as well as around spiral sources.
 The fixed points are encircled in (a), (c), (e), and (g).
  }
  \label{th2}
\end{figure}

\begin{figure}[h]
  \centering
  \includegraphics[width=17.0cm]{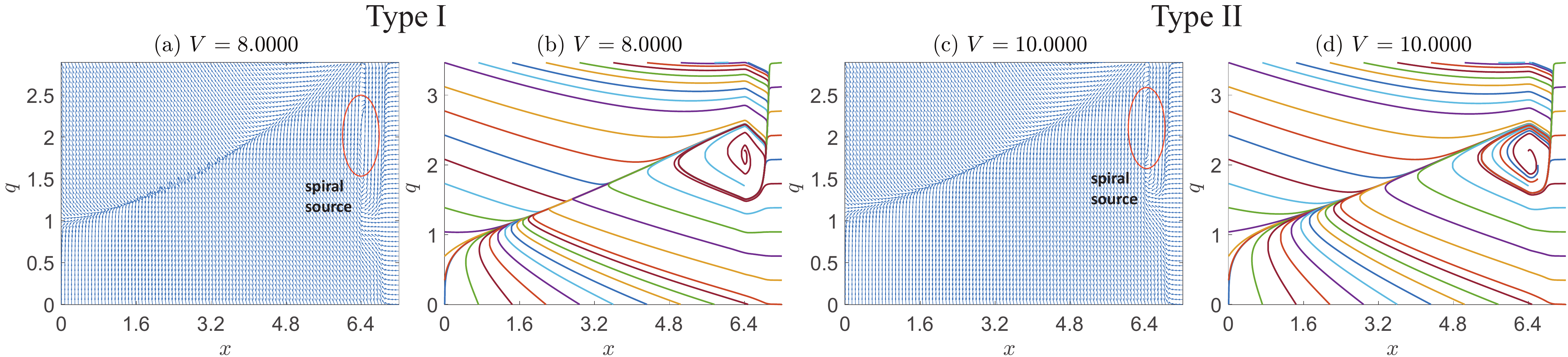}
  \caption{(a), (c) Normalized flow fields and (b), (d) trajectories for $V_1'<V<V_2$.
   To obtain these plots, the initial conditions were set at equal spacing along the edge of the plots. The fixed points are encircled in (a) and (c). Note that in both cases, the only fixed point is a spiral source, around which a limit cycle is a global attractor.
  }
  \label{th3}
\end{figure}

\begin{figure}[h]
  \centering
  \includegraphics[width=17.0cm]{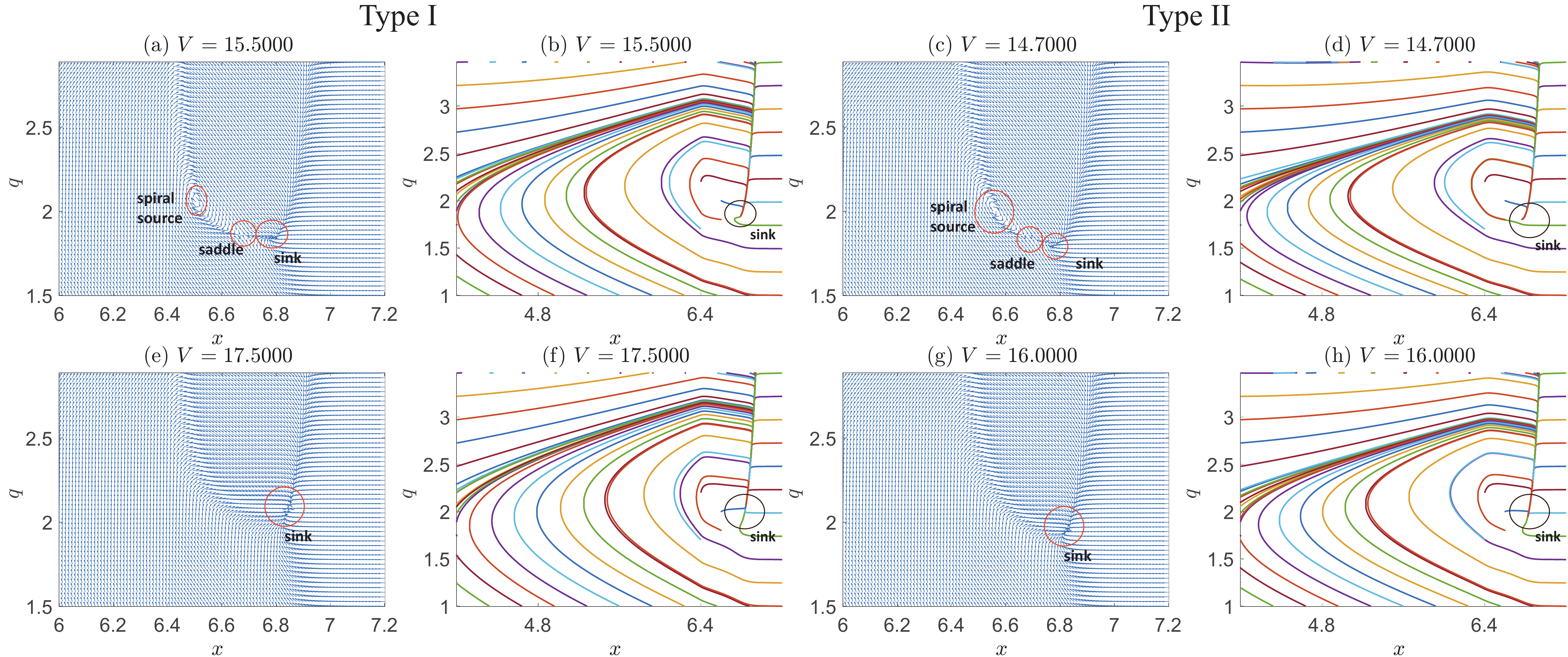}
  \caption{(a), (c), (e), (g) Normalized flow fields and (b), (d), (f), (h) trajectories for $V>V_2$.
  To obtain these plots, the initial conditions were set at equal spacing along the edge of the plots. The fixed points are encircled in (a), (c), (e), and (g).}
  \label{th45}
\end{figure}

The vector fields for several representative cases  are shown in Fig.~\ref{th0}-\ref{th45}, where all fixed points are denoted by circles. We used $0.05$ as the evolution time, which was sufficient for the system to reach the attractor. The plotting range was selected to include all fixed points in the relevant regime. Jacobi matrices  defined as
\begin{equation}
    \mathbf{J}=\frac{\partial\left(g_1,g_2\right)}{\partial\left(x,q\right)}
\end{equation}
were used for the comparison of special points in both models. The functions $g_1(x,q)=\dot{x}$ and $g_2(x,q)=\dot{q}$ were taken from the main equations of the model.
Tables~\ref{tab:Jacb I} and \ref{tab:Jacb II} provide equilibrium points and their characteristics obtained based on the Jacobi matrices. From these comparisons, we conclude that the overall dynamics is the same for the I and II type of memristance.

\begin{table}[]
\centering
\caption{Numerical characteristics of Jacobi matrices $\mathbf{J}$ for fixed points in Fig.~\ref{th0}-\ref{th45} for type I $R(x)$ model. Here, $\Delta=(\text{Tr }\mathbf{J})^2-4\text{det }\mathbf{J}$. }
\label{tab:Jacb I}
\begin{tabular}{rrrrrrc}
\hline
\multicolumn{1}{c}{$V$} & \multicolumn{1}{c}{$x$} & \multicolumn{1}{c}{$q$} & \multicolumn{1}{c}{$\text{det }\mathbf{J}$} & \multicolumn{1}{c}{$\text{Tr }\mathbf{J}$} & \multicolumn{1}{c}{$\Delta$} & Type \\ \hline
0       & 0      & 0      & 53386000    & -14675  & 1799400      & sink          \\
4.0000  & 0.3253 & 0.5207 & 46873000    & -14349  & 18401000     & sink          \\
        & 6.1776 & 2.1927 & -82951000   & 1436    & 333870000    & saddle        \\
        & 6.4011 & 2.1315 & 16939000000 & 15938   & -67502000000 & spiral source \\
6.9000  & 1.2523 & 1.0216 & 28312000    & -13421  & 66869000     & sink          \\
        & 4.3978 & 1.9136 & -34780000   & -10225  & 243670000    & saddle        \\
        & 6.4083 & 2.1277 & 11226000000 & 15265   & -44672000000 & spiral source \\
7.0000  & 1.3120 & 1.0456 & 27119000    & -13361  & 70041000     & sink          \\
        & 4.3078 & 1.8940 & -32957000   & -10324  & 238410000    & saddle        \\
        & 6.4086 & 2.1275 & 11053000000 & 15248   & -43978000000 & spiral source \\
8.0000  & 6.4120 & 2.1257 & 9343300000  & 15105   & -37145000000 & spiral source \\
15.5000 & 6.7918 & 1.8384 & 551970000  & -72552   & 3056000000   & sink          \\
        & 6.6754 & 1.8575 & -232640000  & 29278   & 1787700000   & saddle        \\
        & 6.5041 & 2.0601 & 614580000   & 21641   & -1990000000  & spiral source \\
17.5000 & 6.8444 & 2.0750 & 2044700000  & -251620 & 55133000000  & sink          \\ \hline
\end{tabular}
\end{table}

\begin{table}[]
\centering
\caption{Numerical characteristics of Jacobi matrices $\mathbf{J}$ for fixed points in Fig.~\ref{th0}-\ref{th45} for type II $R(x)$ model. }
\label{tab:Jacb II}
\begin{tabular}{rrrrrrc}
\hline
\multicolumn{1}{c}{$V$} & \multicolumn{1}{c}{$x$} & \multicolumn{1}{c}{$q$} & \multicolumn{1}{c}{$\text{det }\mathbf{J}$} & \multicolumn{1}{c}{$\text{Tr }\mathbf{J}$} & \multicolumn{1}{c}{$\Delta$} & Type \\ \hline
0       & 0       & 0       & 58671000   & -15467  & 4554500      & sink          \\
6.0000  & 0.6503  & 0.7362  & 45666000   & -14817  & 36880000     & sink          \\
        & 6.0838  & 2.1971  & -74046000  & -2558   & 302730000    & saddle        \\
        & 6.3914  & 2.1365  & 1465700000 & 14073   & -5664600000  & spiral source \\
8.0000  & 1.4035  & 1.0815  & 30604000   & -14064  & 75372000     & sink          \\
        & 4.7989  & 1.9980  & -37553000  & -10549  & 261500000    & saddle        \\
        & 6.4149  & 2.1240  & 3814100000 & 15393   & -15020000000 & spiral source \\
9.0000  & 2.2998  & 1.3843  & 12682000   & -13167  & 122640000    & sink          \\
        & 3.6172  & 1.7360  & -13679000  & -11841  & 194930000    & saddle        \\
        & 6.4239  & 2.1188  & 3619200000 & 15831   & -14226000000 & spiral source \\
10.0000 & 6.4339  & 2.1127  & 3137600000 & 16392   & -12282000000 & spiral source \\
14.7000 & 6.7664  & 1.8031  & 245070000 & -28346   & -176810000   & sink (spiral) \\
        & 6.6907  & 1.8389  & -123710000 & 26224   & 1182500000   & saddle        \\
        & 6.55914 & 2.00512 & 295680000  & 28319   & -380770000   & spiral source \\
16.0000 & 6.8230  & 1.9457  & 1305900000 & -159780 & 20307000000  & sink          \\ \hline
\end{tabular}
\end{table}

\begin{figure}[h]
  \centering
  \includegraphics[width=12.0cm]{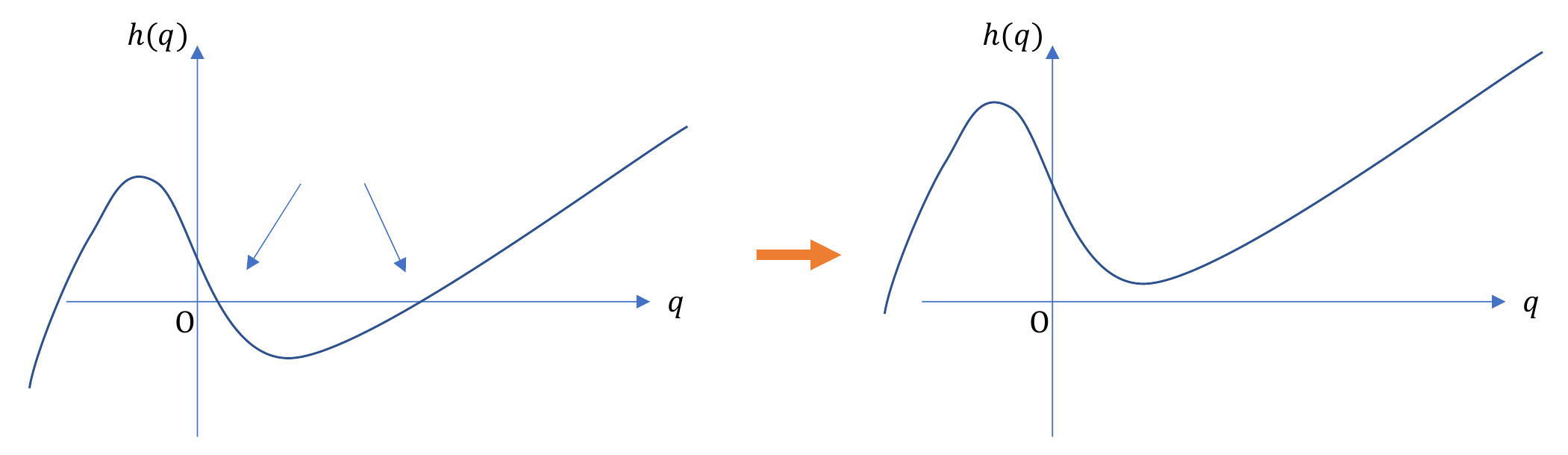}
  \caption{Schematics of $h(q)$. As voltage $V$ increases above $V_1'$, two points of intersection of $h(q)$ with $q$-axis vanish, corresponding to a bifurcation of the sink-saddle pair. }
  \label{h(q)}
\end{figure}

Note the threshold $V_1'$ originates from the bifurcation of the sink-saddle pair located beyond the contact regime. Its approximate location can be  derived analytically if we ignore the Lennard-Jones-like potential and contact drop of resistance. In the normal regime, we have
\begin{equation}
    U(x)\approx\frac12 kx^2  \label{eq:pot_normal} \;\;,
\end{equation}
\begin{equation}
    R(x)\approx\left\{
    \begin{aligned}
        R_m,\quad &\text{for type I memristance model},\\
        10\rho_0\frac{d-x}{A}, \quad &\text{for type II memristance model}.
    \end{aligned}
    \right.
\end{equation}
For type I, by solving the equations analytically, we can define
\begin{equation}
    h(q)\equiv q^3-k\epsilon Adq+\frac{k\epsilon^2 A^2V}{1+r/R_m}\;\;,
\end{equation}
with the fixed-point values of $q$  determined by the condition $h(q)=0$. The schematic behavior of the function $h(q)$ is shown in Fig.~\ref{h(q)}. There are 2 extreme points $q_\pm=\pm\sqrt{k\epsilon Ad/3}$, and $h(q_+)=0$ gives the threshold $V_1'$, i.e.
\begin{equation}
    V_1'=2\left(1+\frac{r}{R_m}\right)\sqrt{\frac{k}{\epsilon A}\left(\frac{d}{3}\right)^3}\;\;,
    \label{r/R}
\end{equation}
and $V_1'=7.9582$. Similarly, for type II it is
\begin{equation}
    V_{1,II}'=2\sqrt{\frac{k}{\epsilon A}\left(\frac{d+r/\rho_0}{3}\right)^3}\;\;
\end{equation}
and $V_{1,II}'=7.1053$. Another ``fixed point'' is nonphysical as it corresponds to $x\geq d$.

\section*{Appendix B: Switching between non-spiking and spiking behavior}

\begin{figure}[h]
  \centering
\includegraphics[width=15.0cm]{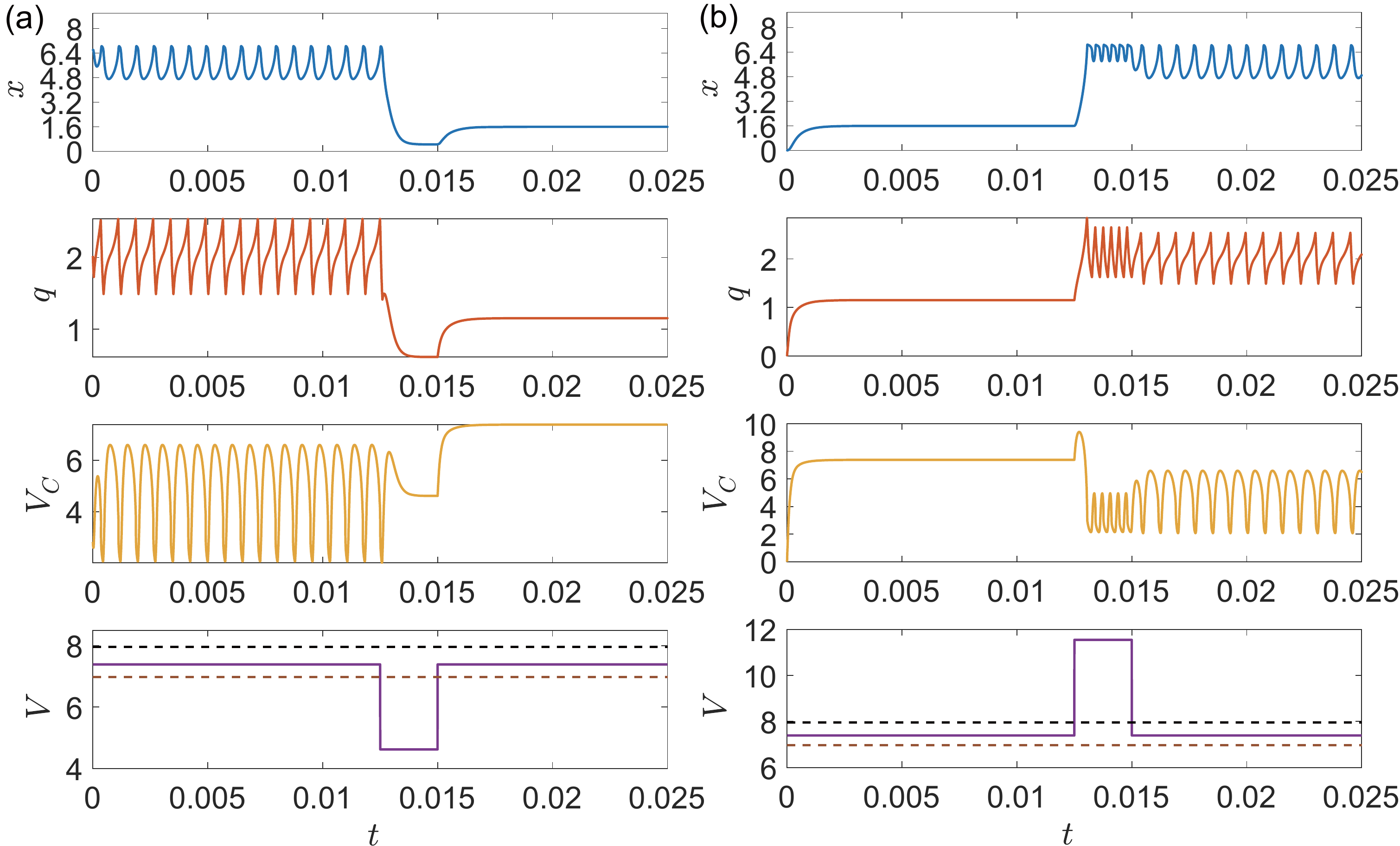}
  \caption{Switching between the static regime and  periodic spiking regime using a DC voltage pulse. (a)~The transition from the periodic spiking state to the static state. This plot was obtained using the initial conditions $x=6.6000$ and $q=2.0207$. (b)~The transition from the static state to the periodic spiking state (the initial conditions were $x=0$ and $q=0$). The black dashed line refers to $V_1'$ while the brown one -- to $V_1$.
  }
  \label{switch}
\end{figure}

Here we show the possibility of switching between the sink and limit cycle. These features co-exist in the phase diagram at $V_1<V<V'_1$.
The transition between these two regimes (in either direction) can be induced by  an appropriate voltage pulse. This is illustrated in Fig.~\ref{switch}.

For the switching from the non-spiking to spiking behavior, the pulse amplitude should exceed  the threshold $V_1'$. For the reverse switching,
the pulse amplitude should be below $V_1$. Of course, in both cases, the pulse should be of appropriate duration.

\section*{Appendix C: Frequency modulation}

\begin{figure}[h]
  \centering
  \includegraphics[width=17.0cm]{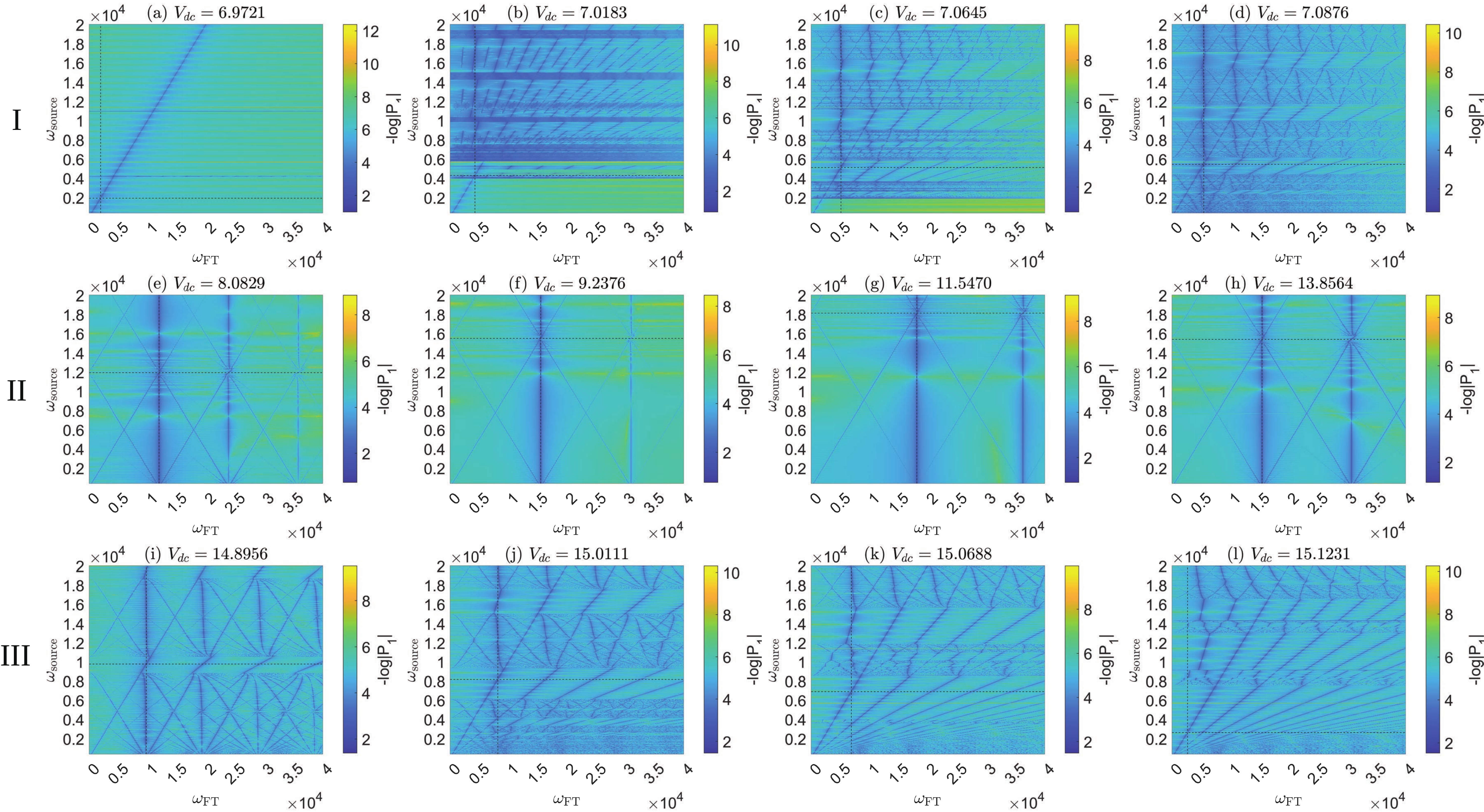}
  \caption{Fourier transforms of $V_C(t)$ in response to external voltage with an extra ac component.
  Different oscillation regimes are represented as follows: panels (a)-(d) represent regime I, panels (e)-(h) correspond to regime II, and
  panels (i)-(l) are related to regime III. In these calculations, to keep the system close to the limit cycle, the initial conditions were selected as $x_0=6.6000$ and $q_0=2.0207$. The dashed lines refer to corresponding natural frequencies. $P_1$ is the single-sided amplitude of the Fourier transform. These plots were obtained using the evolution time $\geq 1$; the initial interval of transient dynamics was omitted.
  }
  \label{FTsynAll}
\end{figure}

\begin{figure}[h]
  \centering
  \includegraphics[width=17.0cm]{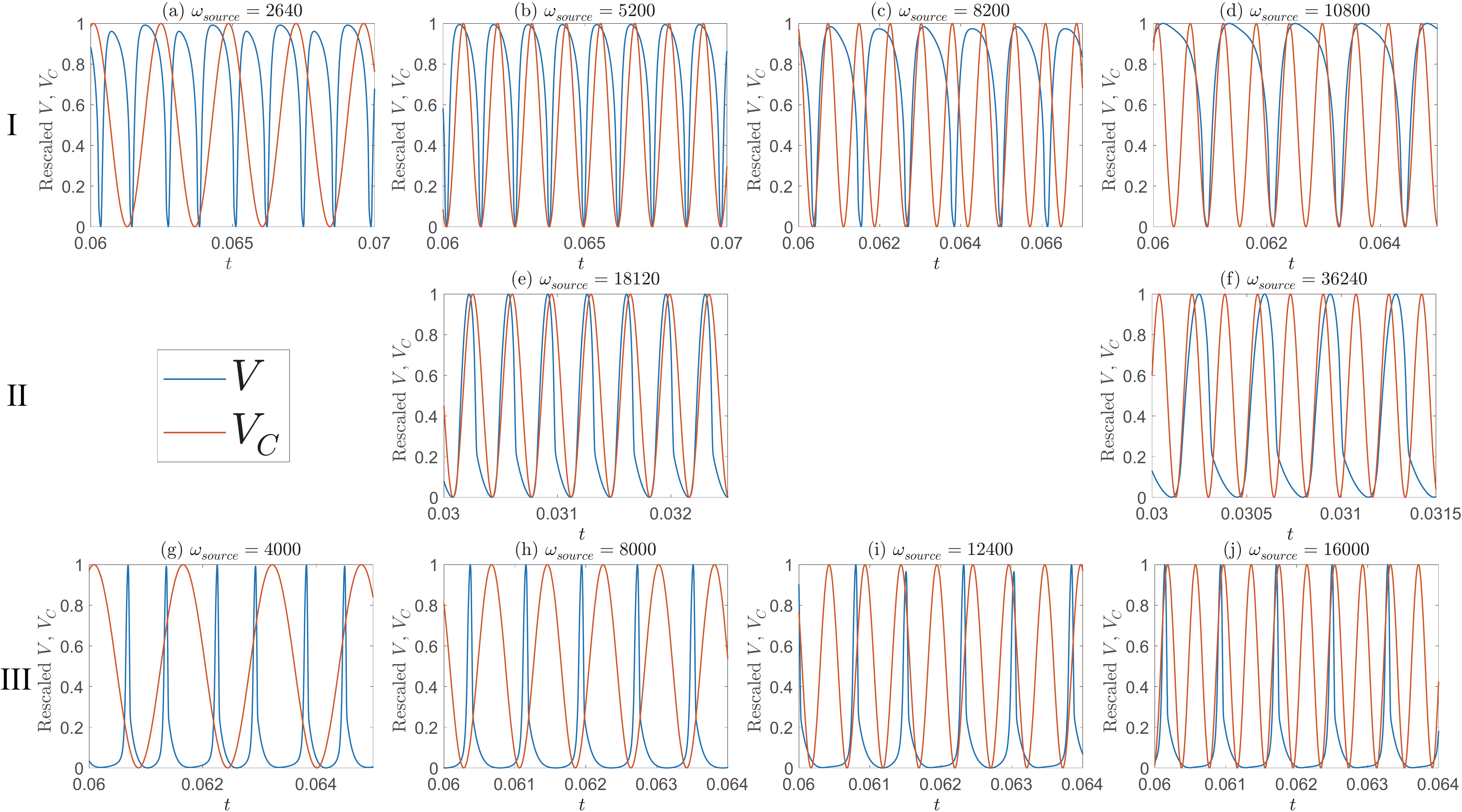}
  \caption{Comparisons of  $V_C(t)$ and $V(t)$ in the synchronized regions of (a)-(d) regime I ($V=7.0876$), (e)-(f) regime II ($V=11.5470$), and (g)-(j) regime III ($V=15.0111$). Both $V_C$ and $V$ are re-scaled between 0 and 1 for convenience. The initial conditions were selected as $x_0=6.6000$ and $q_0=2.0207$.
  }
  \label{synLock}
\end{figure}

More detailed Fourier transforms of the frequency modulation of synchronization on the three regimes are shown in Fig. \ref{FTsynAll}, from which we can capture the evolution of the modulating picture when approaching the thresholds $V_1$ and $V_2$. Furthermore, in Fig. \ref{synLock}, we examine the locking behaviors of $V_C$ in the three regimes. As shown in Fig. \ref{synLock}(a)(g), there are two spikes in one period of $V$, while this number of spikes is one in Fig. \ref{synLock}(b)(e)(h), two-thirds in Fig. \ref{synLock}(c)(i), and one-half in Fig. \ref{synLock}(d)(f)(j).

\end{document}